\documentclass{cjpsuppl}
\newcommand{\lsim}{\raisebox{-0.13cm}{~\shortstack{$<$ \\[-0.07cm] $\sim$}}~}
\newcommand{\gsim}{\raisebox{-0.13cm}{~\shortstack{$>$ \\[-0.07cm] $\sim$}}~}
\newcommand{\nn}{\noindent}

\newcommand{\ra}{\rightarrow}
\newcommand{\tb}{\tan \beta}

\newcommand{\beq}{\begin{eqnarray}}
\newcommand{\eeq}{\end{eqnarray}}

\usepackage{epsfig}

\begin{document}
\title{Phenomenology of SM and SUSY Higgs bosons at the LHC}
\authori{Abdelhak Djouadi}
\addressi{LPMT, Universit\'e de Montpellier, F--34095 Montpellier Cedex 5, 
France.\\
LPT, Universit\'e Paris--Sud \& UMR8627--CNRS, Bt. 210, F--91405 Orsay, France. 
}
\authorii{}    \addressii{}
\authoriii{}   \addressiii{}
\authoriv{}    \addressiv{}
\authorv{}     \addressv{}
\authorvi{}    \addressvi{}
\headtitle{Phenomenology of SM and SUSY Higgs bosons at the LHC}
\headauthor{Abdelhak Djouadi}
\lastevenhead{A. Djouadi:  Phenomenology of SM and SUSY Higgs bosons at the LHC}
\pacs{}
\keywords{}
\refnum{}
\daterec{}
\suppl{A}  \year{2004} \setcounter{page}{1}
\maketitle

\begin{abstract}
I briefly review the physics of the Higgs sector in the Standard Model (SM) and
its supersymmetric extension, in particular the MSSM, and discuss the prospects
for discovering the Higgs particles at the Large Hadron Collider. Some emphasis
will be put on the theoretical developments which occurred in the last few 
years.\smallskip

\centerline{\it Talk given at ``Physics at the LHC", Vienna, July 2004.}
\end{abstract}

\section{A brief introduction}

The search for the Higgs particles \cite{Higgs}, the remnants of the mechanism
introduced forty years ago to achieve the breaking of the electroweak symmetry,
is the primary mission of the LHC.  Detailed theoretical and experimental
studies performed in the last few years, have shown that the single neutral
Higgs boson that is predicted in the SM \cite{HHG} can be detected at the LHC
\cite{LHC,Houches,Houches03} over its entire mass range, 114.4 GeV $\le M_H
\lsim 1$ TeV, in many redundant channels; see Fig.~1 (left).  [It could also be
discovered at the upgraded Tevatron, if it is relatively light and if enough
integrated luminosity is collected; see Ref.~\cite{Tevatron} for instance.] In
the context of the Minimal Supersymmetric Standard Model (MSSM), where the 
Higgs sector is extended to contain two CP--even neutral Higgs bosons $h$ and
$H$, a pseudoscalar $A$ boson and a pair of charged scalar particles $H^\pm$
\cite{HHG}, it has been shown that the lighter $h$ boson cannot escape
detection at the LHC and that in large areas of the parameter space more than
one Higgs particle can be found; Fig.~1 (right).

However, discovering the Higgs boson is not the entire story and another goal,
that is just as important, is to probe the electroweak symmetry breaking
mechanism in all its facets. Once Higgs bosons are found, the next step would
therefore be to explore all their fundamental properties. To achieve this goal
in great detail, one needs to measure all possible cross sections and decay
branching ratios of the Higgs bosons to derive their masses, their total decay
widths, their couplings to the other particles and their self--couplings, their
spin--parity quantum numbers, etc...  This needs very precise theoretical
predictions and more involved and combined theoretical and experimental
studies. In particular, all possible production and decay channels of the Higgs
particles, not only the dominant and widely studied ones allowing for clear
discovery, should be investigated.  

In this talk, I will summarize some studies that were performed recently in the
SM and MSSM Higgs sectors; a few remarks will be made for non--minimal SUSY
extensions. 
After a r\'esum\'e of the present constraints, I discuss the new
developments in the calculation of the Higgs spectrum, the decay and production
rates and possible measurements of the Higgs properties at the LHC.

\begin{figure}[hbtp]
\vspace*{-.6cm}
\caption{The integrated luminosity needed for the  discovery of the SM Higgs
boson  at the LHC in various production and decay channels (left) and the
number of  Higgs particles that can be detected in the MSSM  $[\tb,M_A]$ 
parameter space (right); from Refs.~\cite{LHCplots}.}
\vspace*{-.3cm}
\begin{center}
\epsfig{file=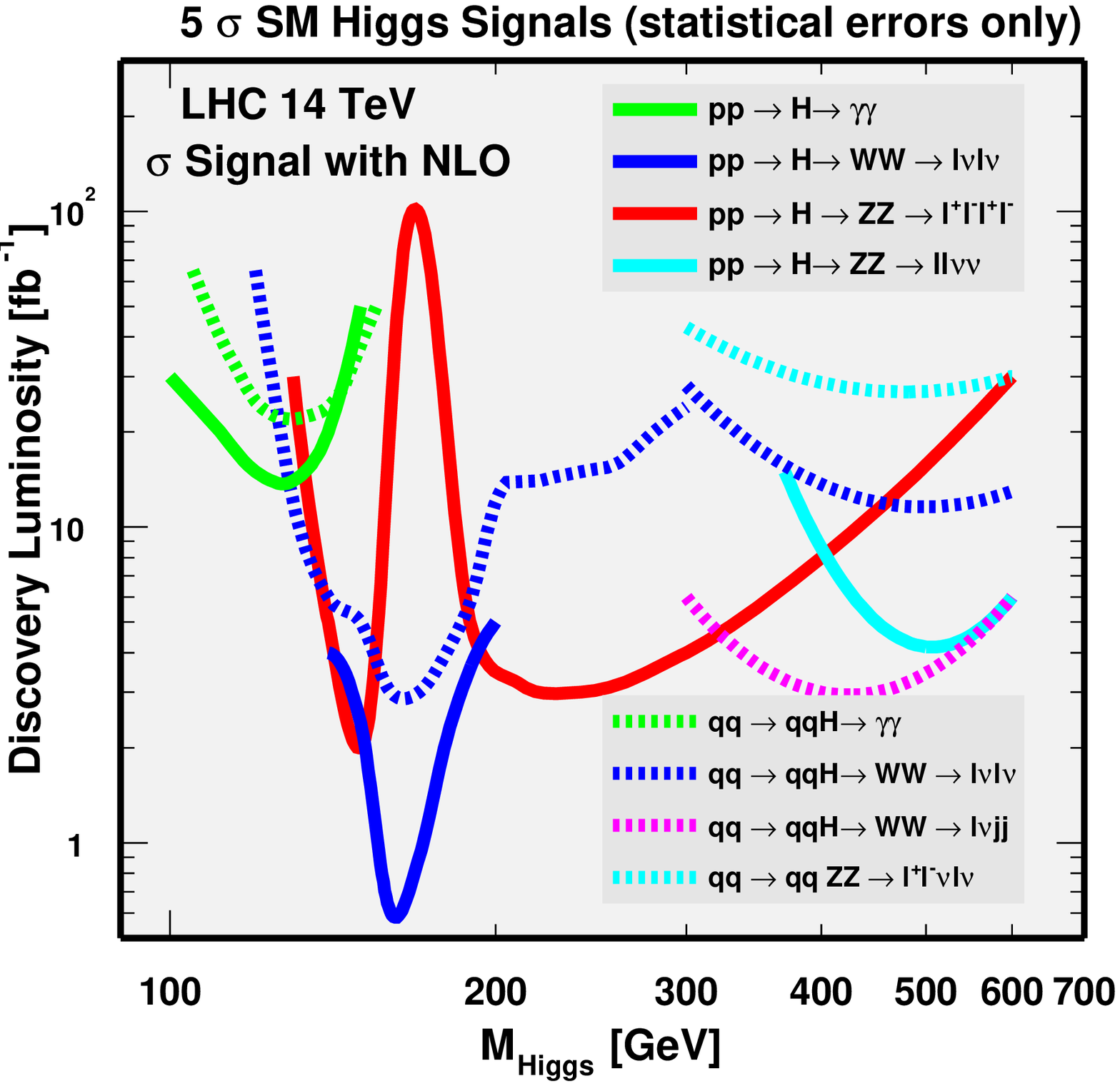,width=6.5cm,height=6.6cm}\hspace*{-3mm}
\epsfig{file=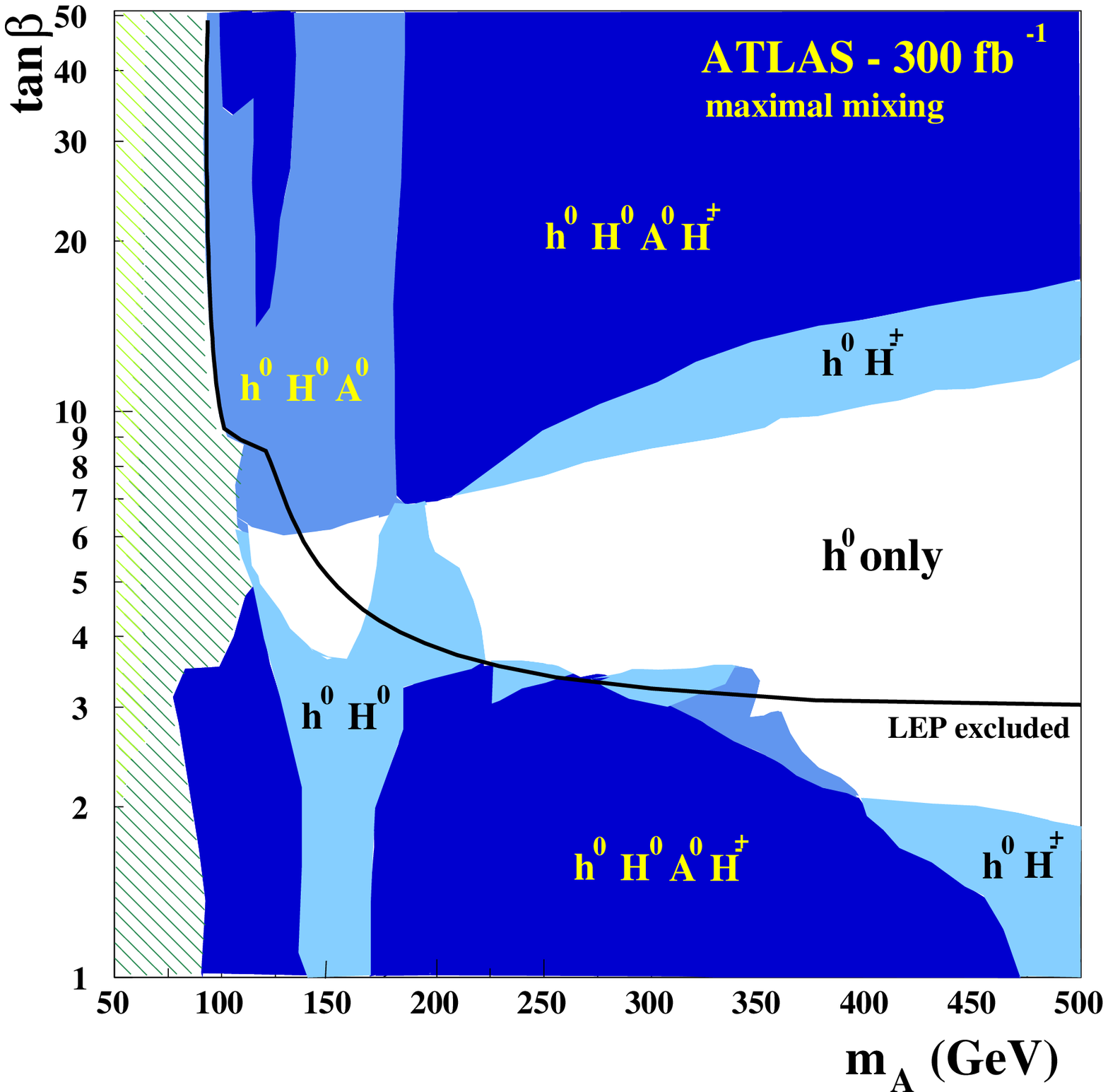,width=6.cm,height=6.1cm}
\end{center}
\vspace*{-10mm}
\end{figure}

\section{The profile of the Higgs bosons}

In the SM, the profile of the Higgs boson is uniquely determined once $M_H$ is
fixed:  the decay width and branching ratios, as well as the production cross
sections, are given by the strength of the couplings to fermions and gauge
bosons, which is set by the masses of these particles. There are two
experimental constraints on this free parameter: $i)$ from the negative
searches for Higgs bosons at LEP2 in the Higgs--strahlung process $e^+e^- \to
HZ$ with c.m. energies up to $\sqrt{s}=209$ GeV, the limit $M_H \geq 114.4$ GeV
is established at the 95\% CL \cite{LEP2} and $ii)$ from the high precision
measurement of electroweak observables at LEP, SLC and the Tevatron and with
the latest Tevatron value of the top quark mass, $m_t=178 \pm 4.3$ GeV
\cite{toptev}, one obtains a preferred Higgs boson mass of $M_H=114^
{+69}_{-45}$ GeV from a global fit of all data, leading to the 95\% CL upper
limit of $M_H <260$ GeV \cite{LEPH}.  

However, theoretical constraints can also be derived from assumptions on the
energy range within which the SM is valid before perturbation theory breaks
down and New Physics should appear.  If $M_H \gsim 1$ TeV, the longitudinal $W$
and $Z$  bosons would interact strongly;  to ensure unitarity in their
scattering at high  energies, one needs $M_H \lsim 710$ GeV at tree--level.  In
addition, the quartic Higgs self--coupling, which at the weak scale is fixed by
$M_H$, grows logarithmically with energy and a cut--off $\Lambda$ should be
imposed before it becomes infinite.  The condition $M_H \lsim \Lambda$ sets an
upper limit at $M_H \sim 630$ GeV, that is confirmed with simulations on the
lattice which take into account the strong interactions near the limit. 
Furthermore, top quark loops tend to drive the coupling to negative values, for
which the vacuum becomes unstable. Requiring the SM to be extended to the GUT
scale, $\Lambda \sim 10^{16}$ GeV, the Higgs mass should lie in the range 130
GeV $\lsim M_H \lsim 180$ GeV. For a review of these issues, see
Ref.~\cite{review} for instance.  

In the MSSM, two doublets of Higgs fields are needed to break the electroweak
symmetry, leading to two CP--even  neutral $h,H$ bosons, a pseudoscalar $A$
boson and a pair of charged  scalar particles, $H^\pm$ \cite{HHG}. Besides the
four masses, two additional  parameters define the properties of the particles:
a mixing angle $\alpha$ in the neutral CP--even sector, and the ratio of the two
vacuum expectation values, $\tb$. Because of Supersymmetry constraints,  only
two of them, e.g.\, $M_A$ and $\tb$, are in fact independent at tree--level.
While the lightest Higgs mass is bounded by $M_h \leq M_Z$, the masses of the
$A,H$ and $H^\pm$ states are expected to be below ${\cal O}(1$ TeV). However,
mainly because of the heaviness of the top quark, radiative corrections are
very important: the leading part  grows as the fourth power of $m_t$ and
logarithmically with the common top squark mass $M_S$; the stop trilinear 
coupling $A_t$ also plays an important role and maximizes the correction for
the value $A_t \sim \sqrt{6} M_S$ \cite{RCreview}. 

Recently, new calculations of the two--loop radiative corrections have been
performed. Besides the already known  ${\cal O}(\alpha_t \alpha_s)$ correction
\cite{RCreview}, the contributions at ${\cal O} (\alpha_t^2)$, ${\cal O}
(\alpha_s \alpha_b)$ and ${\cal O}  (\alpha_b^2)$ have been derived
\cite{Pietro}. [The small ${\cal O}  (\alpha_\tau^2)$ corrections have been
also evaluated \cite{adkps}, which completes the two--loop corrections to the
Higgs masses due to the strong and third--generation Yukawa couplings.] These
corrections are sizable, increasing the predicted value for $M_h$ [for given
$\tb$ and $M_A$ inputs] by several GeV.  Also recently, these radiative
corrections have been implemented \cite{adkps} in three codes for the
determination of the MSSM particle spectrum, which appeared in the last few
years: {\tt SuSpect}, {\tt Softsusy} and {\tt Spheno} \cite{codes}.  

\begin{tabular}{ll} 
\hspace*{-.6cm}
\begin{minipage}{6cm}
\vspace*{-1.1cm}
\nn {\small Fig.~2: The lighter Higgs boson mass $M_h$ in the MSSM as a 
function of $\tan \beta$ as obtained from a full scan of the parameter space for
the top mass values $m_t=$ 173.7, 178.0 and 182.3 GeV. The thick dotted 
line on the top is for the conservative case, where $m_t=182.3$ GeV is used
and a theoretical error [due to the renormalization scale variation and 
scheme dependence, and to the approximation of zero--momentum
transfer in the two--loop contributions] is added linearly; from 
Ref.~\cite{adkps}.}
\end{minipage}
&
\begin{minipage}{14cm}
\vspace*{-.3cm}
\hspace*{-1.2cm}
\epsfig{figure=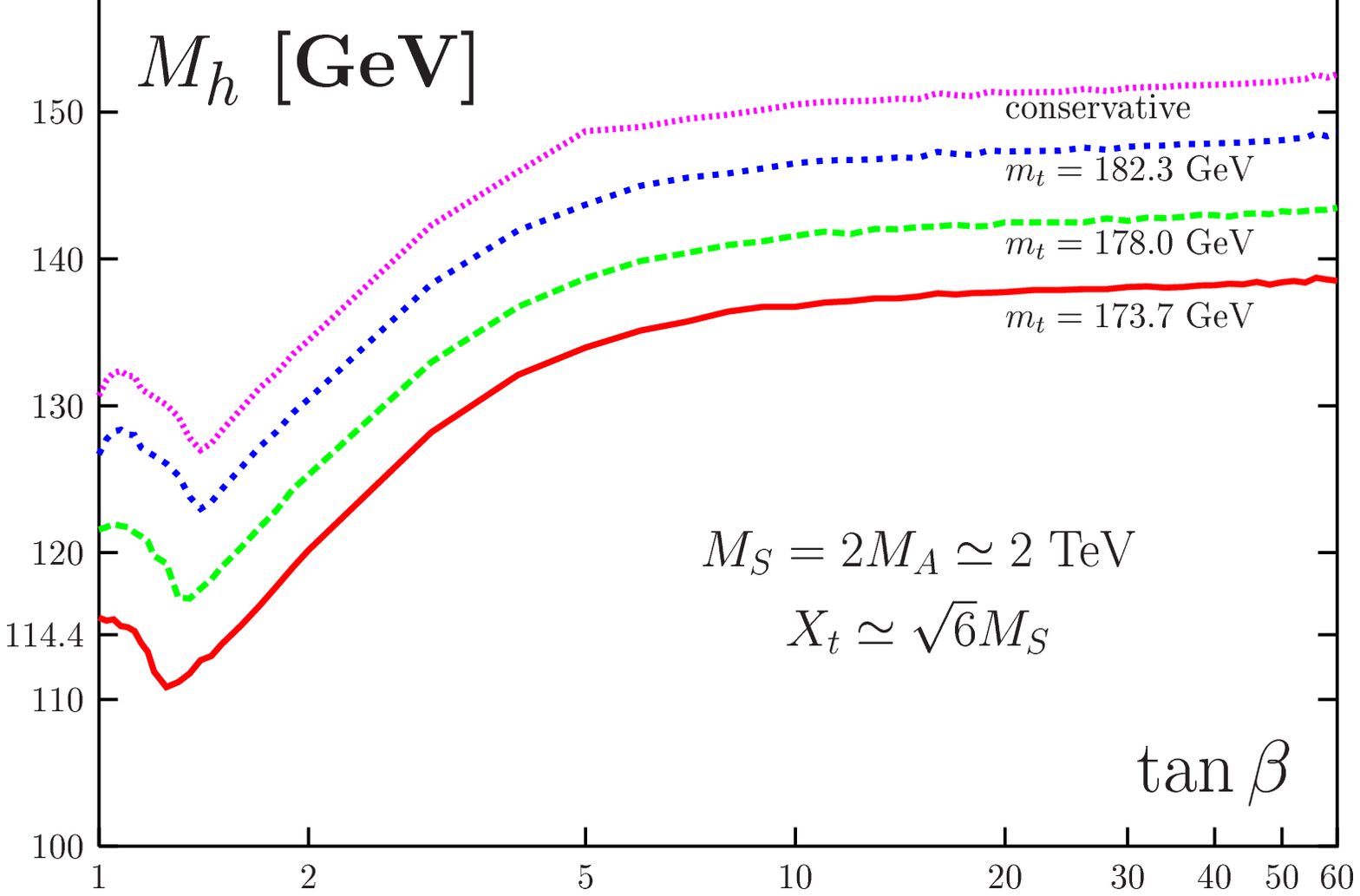,width=8cm,height=14cm}
\vspace*{-7.cm}
\end{minipage}
\end{tabular}
\setcounter{figure}{2}
\vspace*{-1.2cm}

These radiative correction shift the upper bound of the lighter $h$ boson in
the MSSM from the tree--level value, $M_h =M_Z$ by several ten GeV. A full scan
of the MSSM parameter space performed in Ref.~\cite{adkps} shows that the most
conservative upper bound, when $m_t=182.3$ GeV is used and an estimated
theoretical error of 4 GeV is added linearly, is $M_h^{\rm max}\simeq 150$ GeV;
Fig.~2. Furthermore, if one takes into account the absolute limits $M_h \sim
M_A \gsim 92$ GeV from the negative searches at LEP2 \cite{LEP2-MSSM}, as well
as the constraint $M_h \gsim 114$ GeV when the $h$ boson has SM--like
couplings, one can in principle constrain $\tb$. However, as can be seen in
Fig.~2, no lower bound can be set in the conservative case mentioned above, 
provided that $\tb$ is larger than unity which is the case in SUSY extensions 
of the SM.  

The production and the decays of the MSSM Higgs bosons depend strongly on their
couplings to gauge bosons and fermions.  The pseudoscalar has no tree level
couplings to gauge bosons, and its couplings to down\,(up)--type fermions are
(inversely) proportional to $\tb$. It is also the case for the couplings of the
charged Higgs particle to fermions, which are a mixture of scalar and
pseudoscalar currents and depend only on $\tb$. For the CP--even Higgs bosons,
the couplings to down--(up)--type fermions are enhanced (suppressed) with
respect to the SM Higgs couplings for $\tb >1$. They share the SM Higgs
couplings to vector bosons since they are suppressed by $\sin (\beta-\alpha)$
and $\cos(\beta-\alpha)$ factors, respectively for $h$ and $H$.  If the
pseudoscalar mass is large, the $h$ boson mass reaches its upper limit [which
depends on the value of $\tb$] and the angle $\alpha$ reaches the value
$\alpha= \beta-\frac{1}{2}\pi$. The $h$  couplings to fermions and gauge bosons
are then SM--like, while the heavier CP--even $H$ and charged $H^\pm$ bosons
become degenerate in mass with $A$. In this decoupling limit, it is very
difficult to distinguish the SM and MSSM Higgs  sectors.  

Let us now discuss the Higgs decay modes and branching ratios (BR)
\cite{decays} and start with the SM  case. In the ``low--mass" range,
$M_H \lsim 130$ GeV,  the Higgs boson decays into  a large variety of channels.
The main mode is by far the decay into  $b\bar{b}$ with BR\,$\sim$ 90\%
followed by the decays into $c\bar{c}$ and  $\tau^+\tau^-$ with BRs\,$\sim$
5\%. Also of significance is the top--loop  mediated decay into gluons, which
occurs at the level of $\sim$ 5\%.  The top and $W$--loop mediated
$\gamma\gamma$ and $Z \gamma$ decay modes, which lead to  clear signals, are 
very rare with BRs of ${\cal O}(10^{-3})$.  In the ``high--mass" range, $M_H
\gsim 130$ GeV, the Higgs bosons decay into $WW$ and $ZZ$ pairs, one of the
gauge bosons being possibly virtual  below the thresholds. Above the $ZZ$
threshold, the BRs are 2/3 for $WW$ and  1/3 for $ZZ$ decays, and the opening
of the $t\bar{t}$ channel for higher $M_H$ does not alter  this pattern
significantly.  In the low--mass range, the Higgs is very narrow, with
$\Gamma_H<10$ MeV, but this width becomes wider rapidly,   reaching 1 GeV at
the $ZZ$ threshold. For  very large masses, the Higgs  becomes obese, since
$\Gamma_H \sim M_H$, and can hardly be considered as a resonance. The BRs
and total decay widths are summarized in Fig.~3, which is obtained from a 
recently updated version of the code {\tt HDECAY} \cite{hdecay} and where the 
new value $m_t=178$ GeV is used as input. 

\begin{figure}[!h]
\vspace*{-.9cm}
\caption{The decay branching ratios (left) and the total decay width (right) 
of the SM Higgs boson as a function of its mass, as obtained with an 
updated version of {\tt HDECAY} \cite{hdecay}.} 
\begin{center}
\vspace*{-1.9cm}
\hspace*{-1.cm}
\mbox{
\epsfig{file=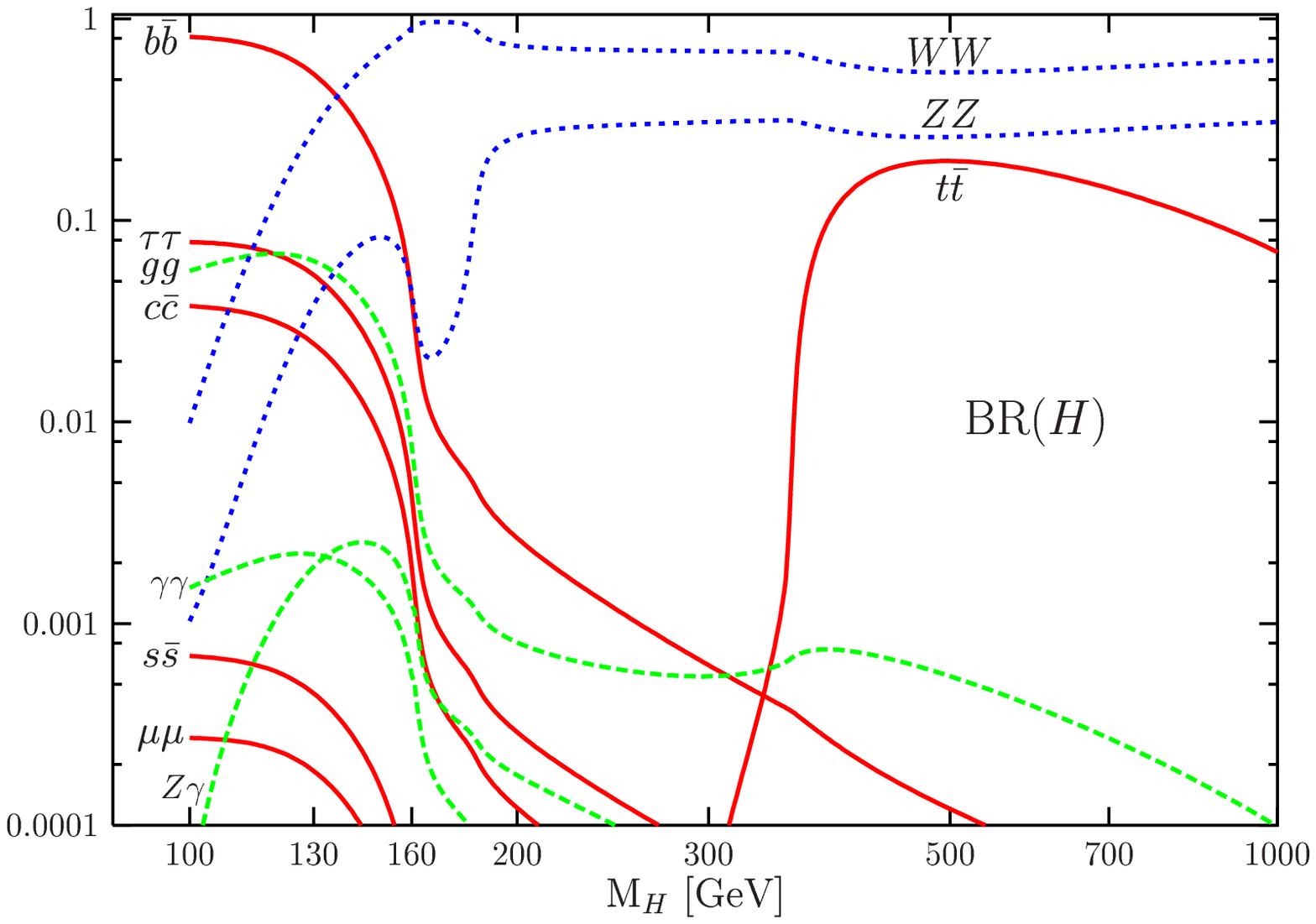,width=7.5cm,height=14cm} \hspace*{-13mm} 
\epsfig{file=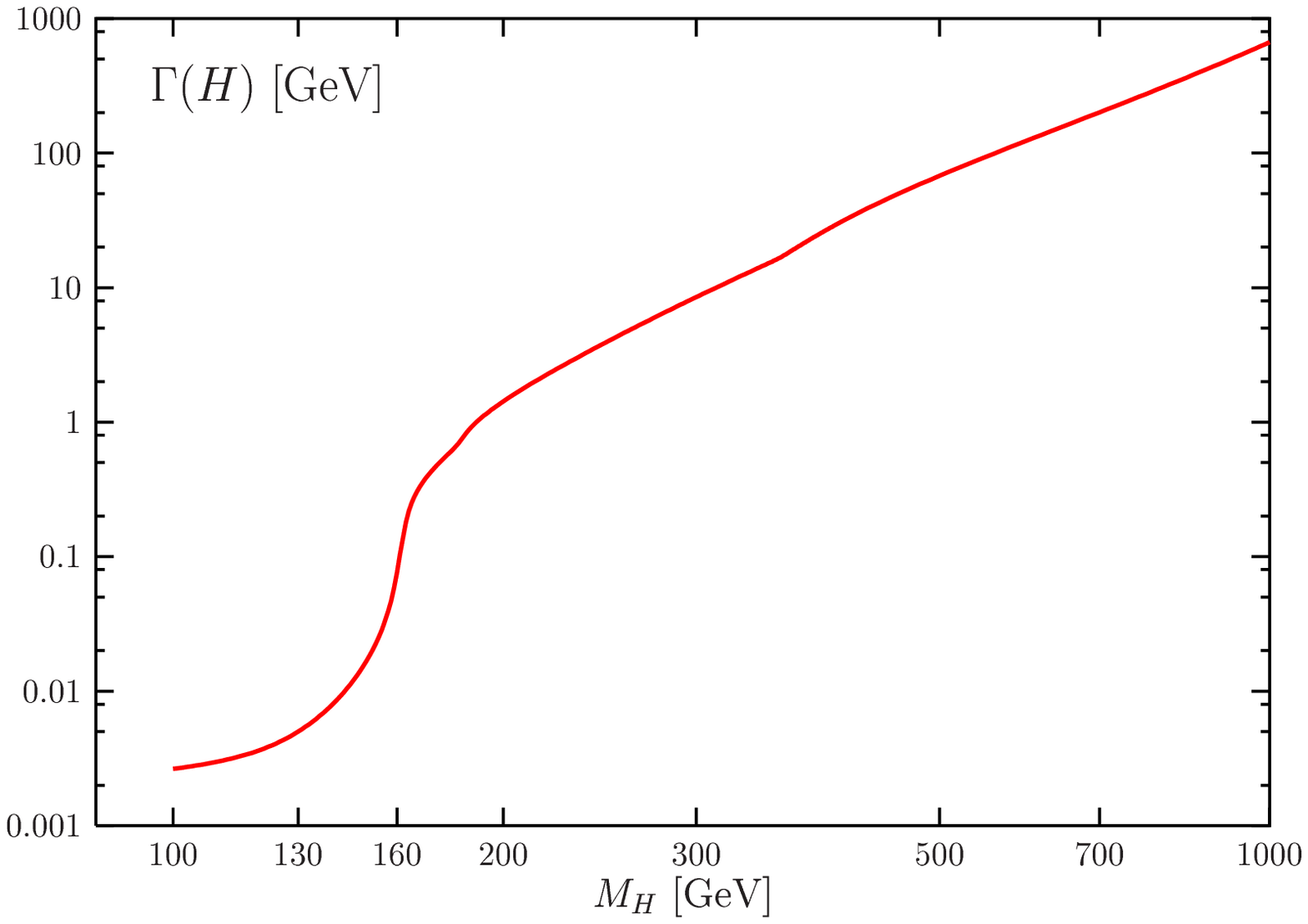,width=7.5cm,height=14cm} 
}
\end{center}
\vspace*{-8.cm}
\end{figure}

In the MSSM, the lightest $h$ boson will decay  mainly into fermion pairs since
$M_h \lsim$~140~GeV; Fig.~4.  This is, in general, also the dominant decay mode
of the $A$ and $H$ bosons, since for $\tb \gg 1$,  they decay into $b \bar{b}$
and $\tau^+ \tau^-$ pairs with BRs of the order of $\sim$ 90\% and 10\%,
respectively. For large masses, the top decay channels $H, A \rightarrow
t\bar{t}$ open up, yet they are suppressed  for large $\tb$. The $H$ boson can
decay into gauge bosons or $h$ boson pairs, and the $A$ particle into $hZ$
final states; however, these decays are strongly suppressed for $\tb \gsim 5$. 
The $H^\pm$ particles decay into fermions pairs: mainly $t\bar{b}$ and $\tau
\nu_{\tau}$ final states for $H^\pm$ masses, respectively, above and below the
$tb$ threshold.  If allowed kinematically, they can also decay  into $hW$ final
states for $\tb \lsim 5$. Adding up the various decays, the widths of all five
Higgsses remain  rather narrow; Fig.~4. Other possible decay channels for the
heavy $H, A$ and $H^\pm$ states, are decays into light charginos and
neutralinos, which could be important if  not dominant
\cite{SUSYdecays,SUSY-Filip}; decays of the $h$ boson into the lightest
neutralinos (LSP) can also be important, exceeding 50\% in some parts of the
parameter space and altering the searches at hadron colliders. Light SUSY
particles can also affect the BRs of the loop--induced modes in a sizable way
\cite{SUSYloops}.  

\begin{figure}[!h]
\vspace*{-.9cm}
\caption{The decay branching ratios and total widths of the MSSM 
Higgs bosons as functions of their masses for $\tb=3$ and 30 as obtained with 
an update of  {\tt HDECAY}; $m_t=178$ GeV and the maximal mixing 
scenario $X_t=\sqrt{6}M_S$ with $M_S=2$ TeV are assumed.} 
\begin{center}
\vspace*{-1.5cm}
\hspace*{-1.2cm}
\mbox{
\epsfig{file=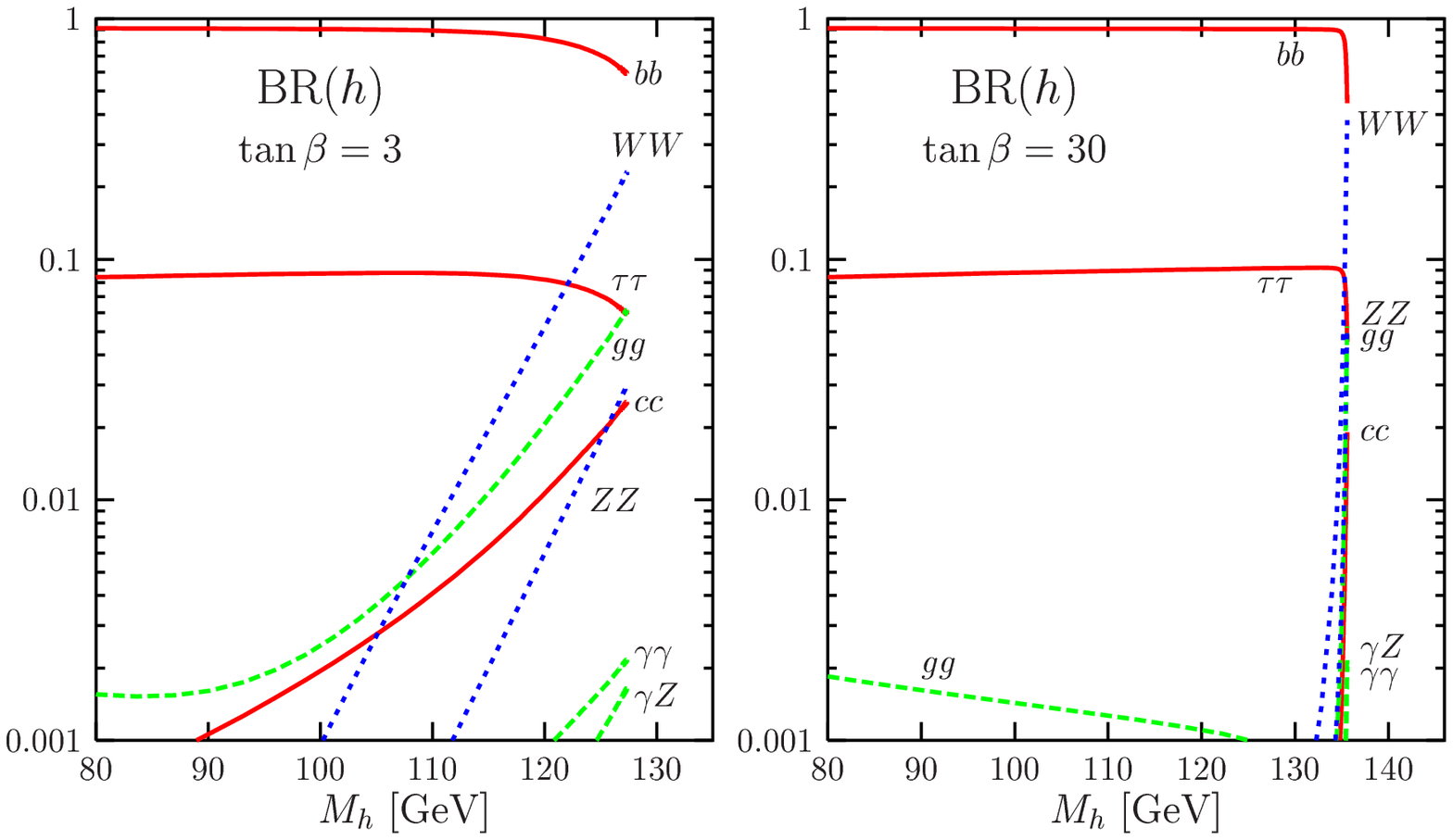,width=7.2cm,height=10cm} \hspace*{-8mm} 
\epsfig{file=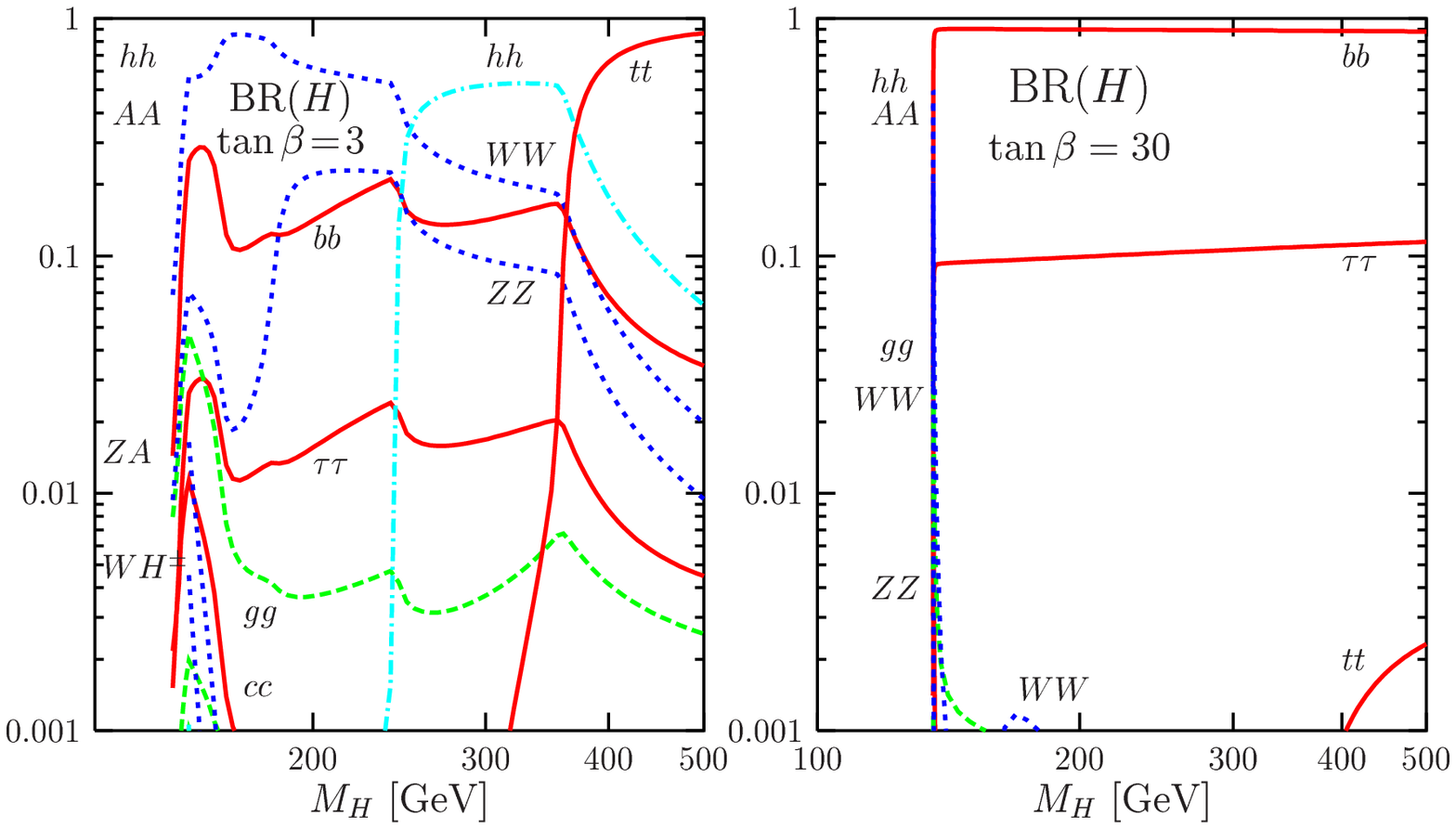,width=7.2cm,height=10cm} 
}
\end{center}
\vspace*{-7.1cm}
\end{figure}

\begin{figure}[!h]
\begin{center}
\vspace*{-.7cm}
\hspace*{-1.2cm}
\mbox{
\epsfig{file=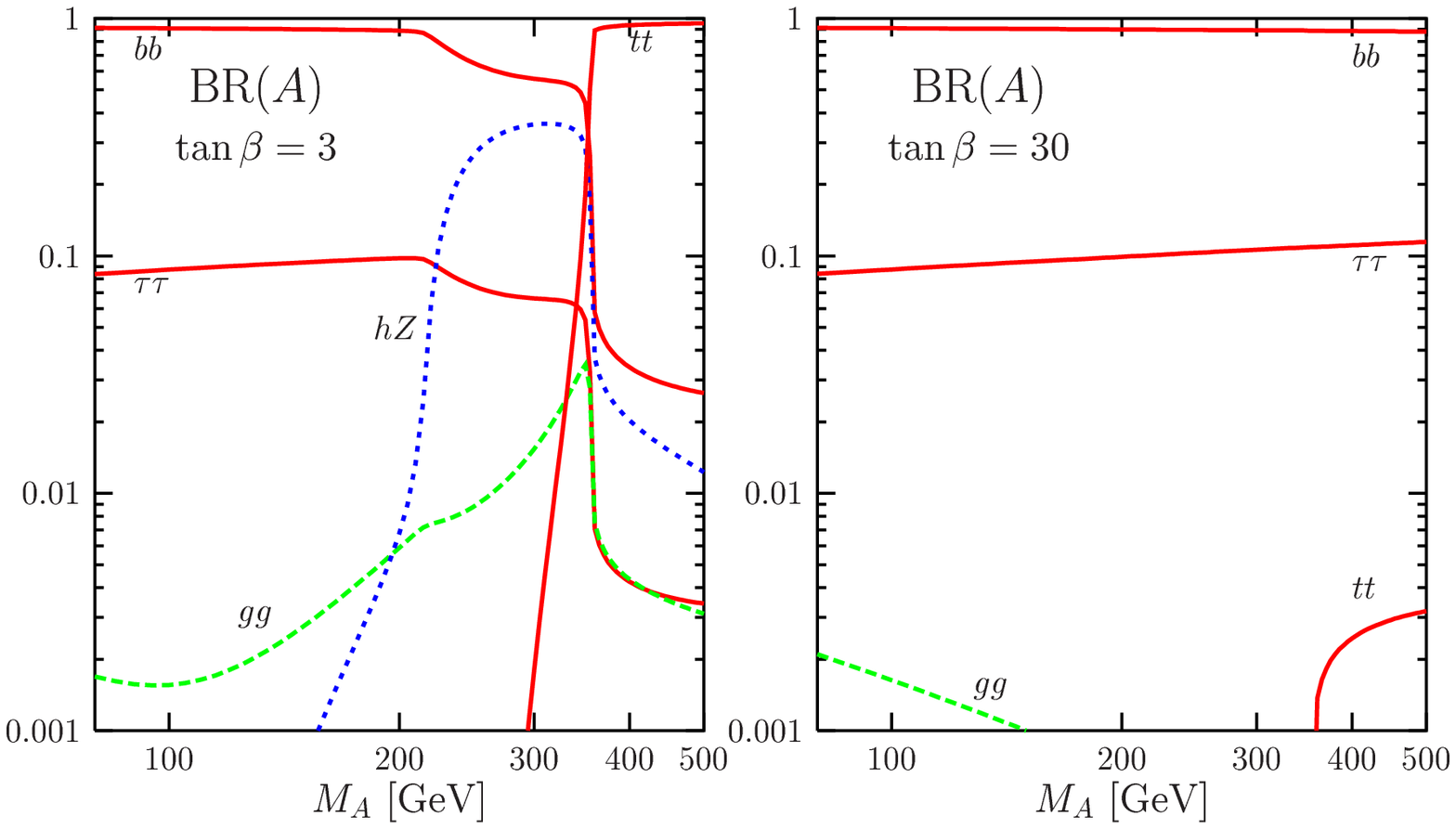,width=7.2cm,height=10cm} \hspace*{-8mm} 
\epsfig{file=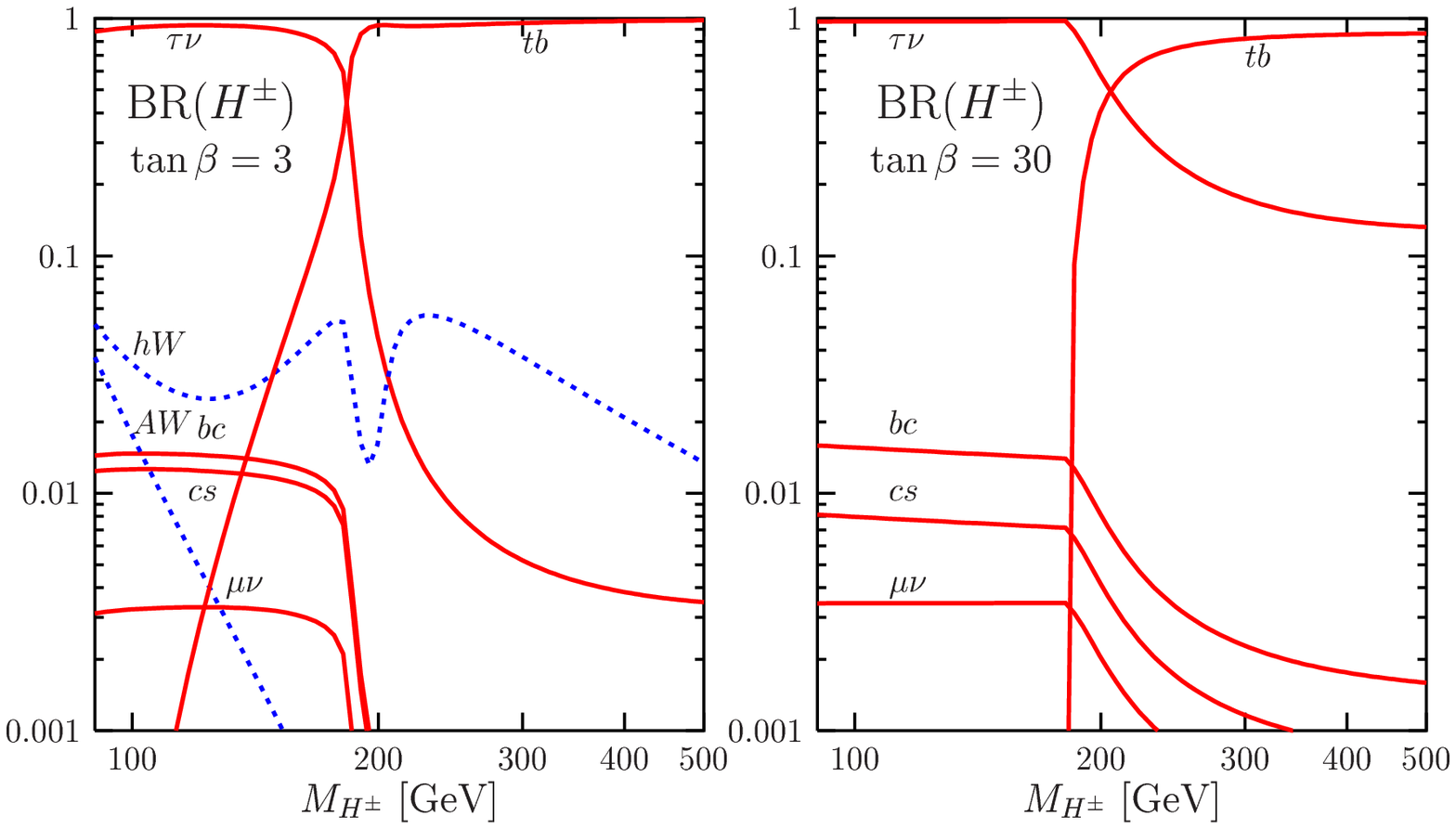,width=7.2cm,height=10cm} 
}
\end{center}
\vspace*{-7.1cm}
\end{figure}

\begin{figure}[!h]
\begin{center}
\vspace*{-.7cm}
\hspace*{-1.2cm}
\mbox{
\epsfig{file=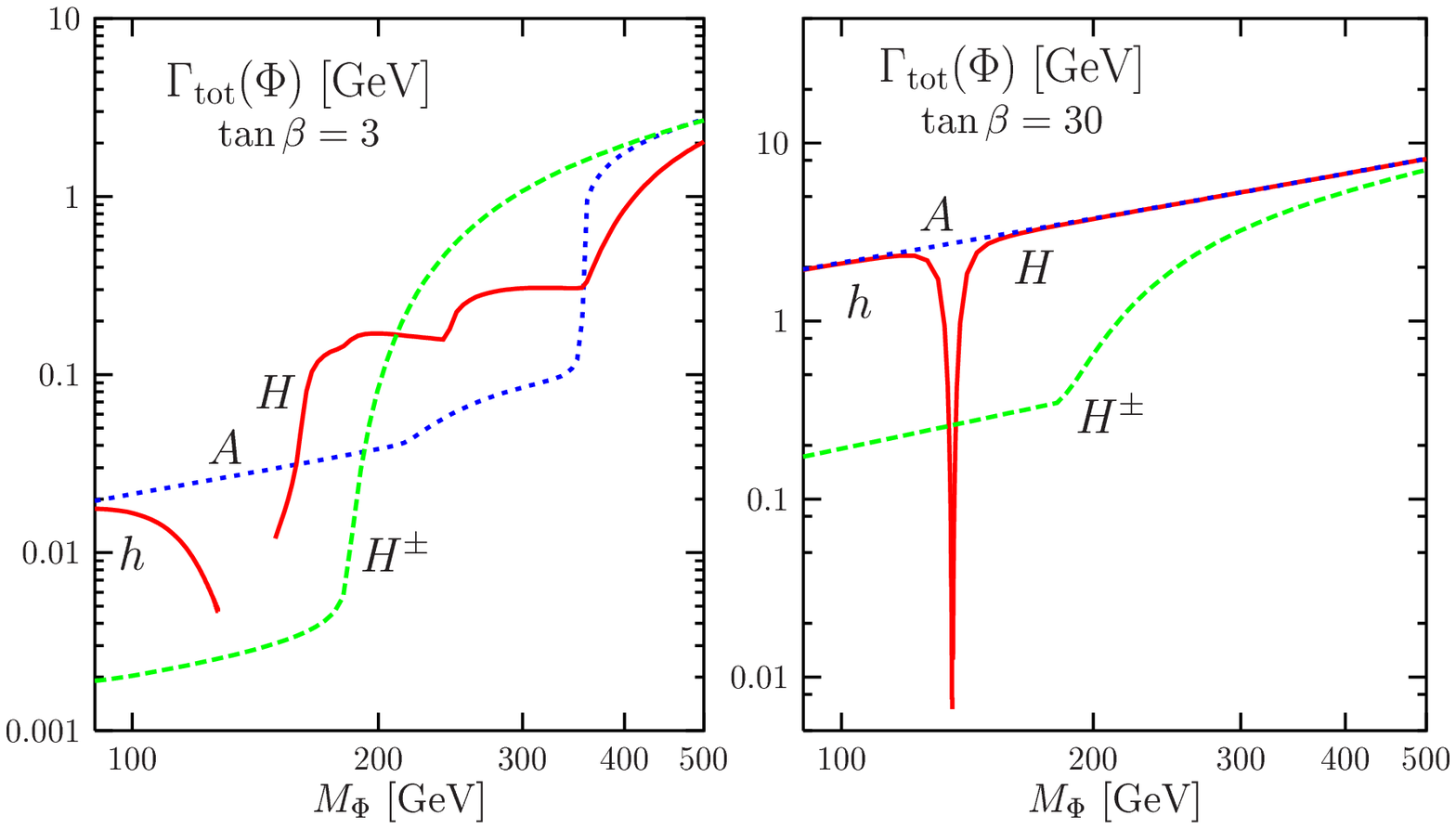,width=7.2cm,height=10cm} 
}
\end{center}
\vspace*{-7.cm}
\end{figure}

\section{SM Higgs production and detection at the LHC}

There are essentially four mechanisms for the single production of the SM Higgs
boson at hadron colliders \cite{P1}; some Feynman diagrams are shown in Fig.~5.
The total cross sections, obtained with the programs of Ref.~\cite{Michael}, 
are displayed in Fig.~6 for the LHC with $\sqrt{s}=14$ TeV as a function of the
Higgs mass; the top quark mass is set to $m_t=178$ GeV and the MRST parton
distributions functions have been adopted. The NLO, and eventually NNLO,
corrections have been implemented as will be summarized later.  In the
following, we discuss the main features of each production channel and
highlight the new theoretical developments which occurred in the evaluation of
the cross sections and detection signals at the LHC.

\begin{figure}[!h]
\vspace*{-.8cm}
\caption{The production mechanisms for SM Higgs bosons at hadron colliders.}
\begin{center}
\vspace*{-1.cm}
\hspace*{-1.5cm}
\epsfig{file=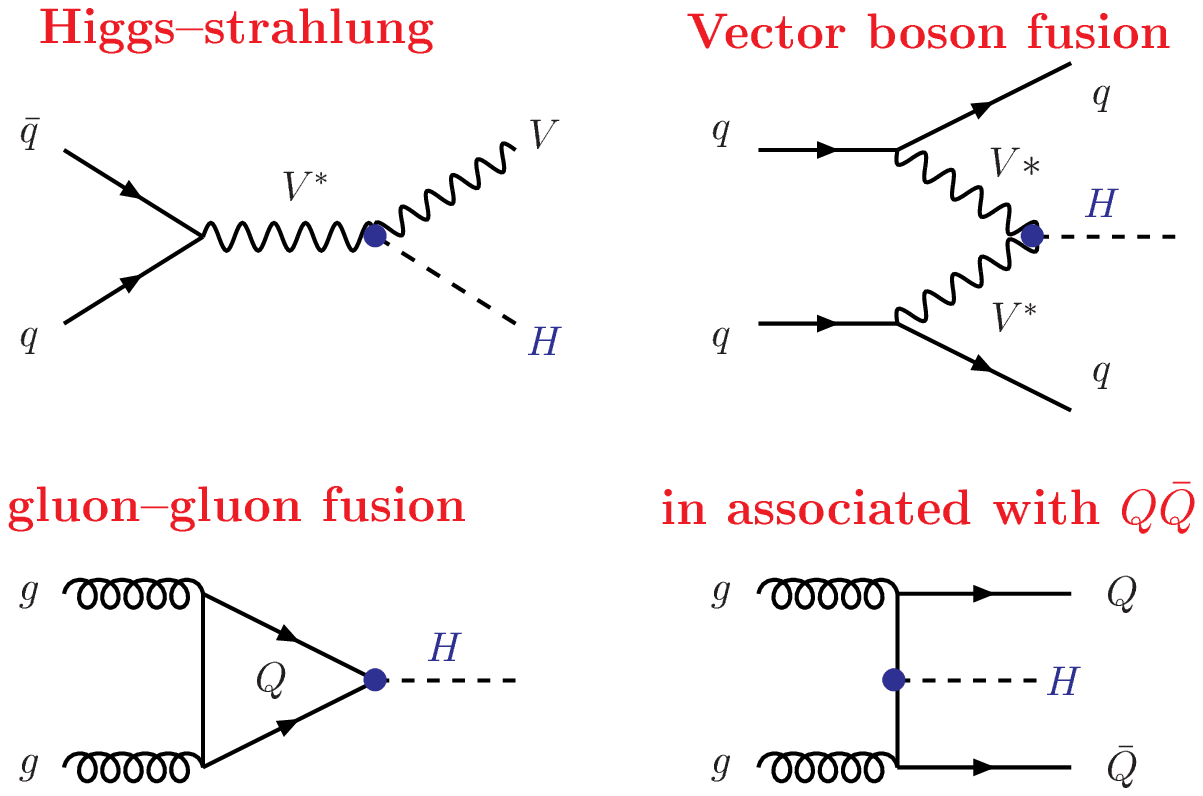,width=12.cm,height=15cm}
\end{center}
\vspace*{-10.7cm}
\end{figure}

\begin{figure}[!h]
\vspace*{-.4cm}
\caption{The production cross sections for the SM Higgs boson at the LHC.}
\begin{center}
\vspace*{-.4cm}
\hspace*{-1.cm}
\epsfig{file=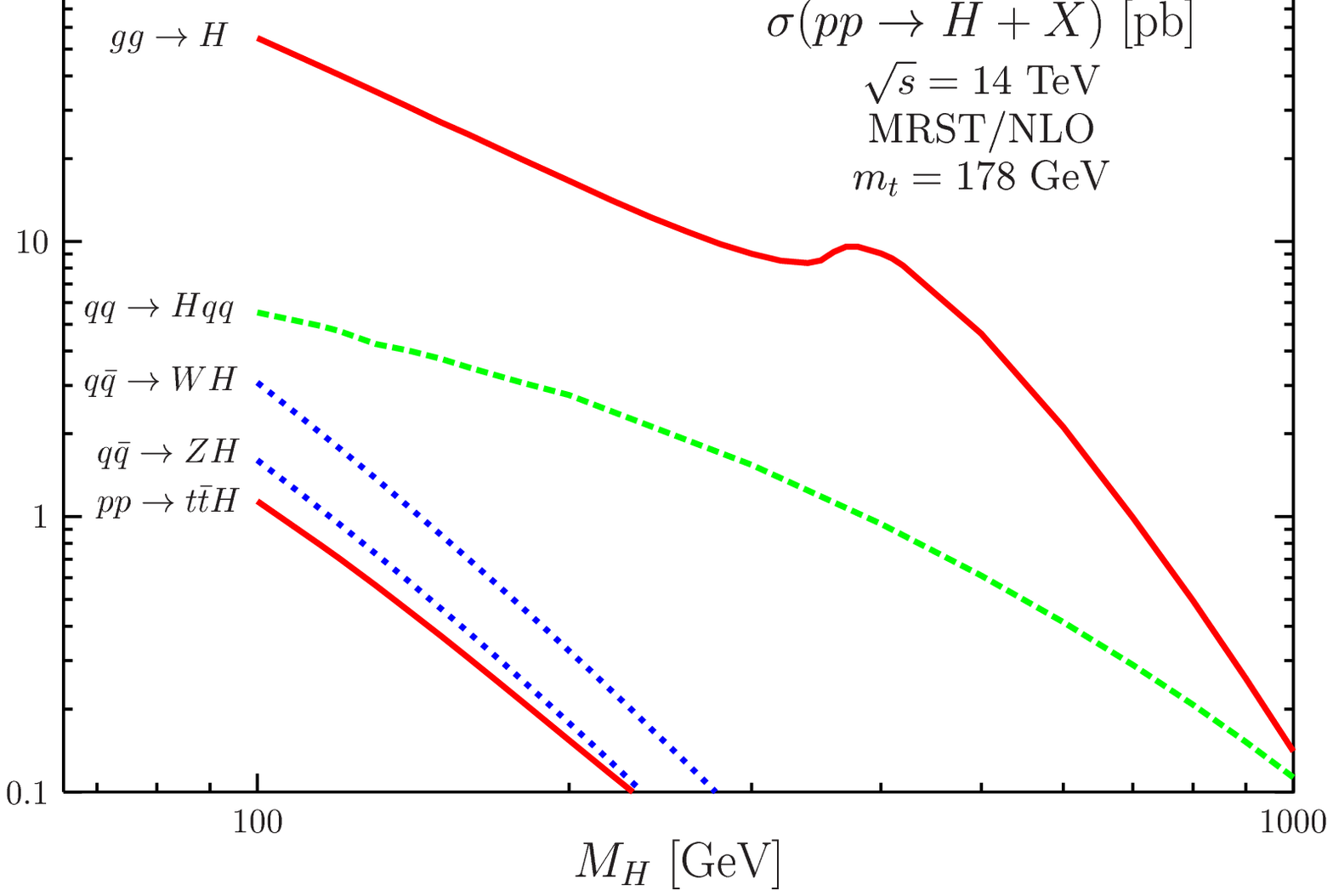,width=11.cm,height=14.5cm} 
\end{center}
\vspace*{-9.cm}
\end{figure}

{\bf a)} \underline{$gg \to H$}: which is by far the dominant production
process at the LHC, up to masses $M_H \approx 1$ TeV. The most promising
detection channels are \cite{gg-detection} $H \to \gamma \gamma$ for $M_H \lsim
130$ GeV and slightly above this mass value, $H\to ZZ^* \to 4\ell^\pm$ and
$H\to WW^{(*)}\to \ell \ell \nu \nu$ with $\ell=e,\mu$ for masses below,
respectively, $2M_W$ and $2M_Z$. For higher Higgs masses, $M_H \gsim 2M_Z$, it
is the golden mode $H \ra ZZ \ra 4\ell^\pm$, which from $M_H \gsim 500$ GeV can
be complemented by $H \to ZZ \to \nu\bar{\nu} \ell^+ \ell^-$ and $H \to WW \to
\nu \ell jj$ to increase the statistics; see Ref.~\cite{LHC}.  

The next--to--leading order (NLO) QCD corrections have been calculated in both
the limit where the internal top quark has been integrated out \cite{P2inf}, an
approximation which should be valid in the Higgs mass range $M_H \lsim 300$
GeV, and in the case where the full quark mass dependence has  been taken into
account \cite{P2}. The corrections lead to an increase of the  cross sections
by a factor of $\sim 1.7$. Recently, the ``tour de force" of deriving the
three--loop corrections has been preformed in the infinite top--quark mass
limit; these NNLO corrections lead to the increase of the rate by an additional
30\% \cite{P3}. This results in a nice convergence of the perturbative series
and a strong reduction  of the scale uncertainty, which is the measure of
unknown higher order effects; see  Fig.~7 (left). The resummation of the soft
and collinear corrections, performed at next--to--next--to--leading logarithm
accuracy,  leads to another increase of the rate by $\sim 5\%$ and a decrease
of the scale uncertainty \cite{SG-resum}.  The QCD corrections to the
differential  distributions, and in particular to the Higgs transverse momentum
and rapidity distributions, have also been recently calculated at NLO [with a
resummation for the former] and shown to be rather large \cite{Pt-eta-distrib}.
The dominant components of the electroweak corrections, some of which have been
derived very recently, are comparatively very  small \cite{EW-CR}.  
 
\begin{figure}
\vspace*{-9mm}
\caption{Left: SM Higgs boson production cross sections in the $gg$ fusion 
process at the LHC as a function  of $M_H$ at the three different orders
with the upper (lower) curves  are for the choice of the 
renormalization and factorization scales $\mu=\frac{1} {2} M_H$ ($2M_H$); 
from Harlander and Kilgore in Ref.~\cite{P3}. Right: $K$-factors for $pp \to 
HW$ at the LHC as a function of $M_H$ at LO, NLO and NNLO with the bands 
represent the spread of the cross section when the scales are varied in the 
range $\frac{1}{3} M_{HV} \leq \mu_R\, (\mu_F) \leq 3M_{HV}$; from 
Brein et al. Ref.~\cite{HV-NNLO}.}
\vspace*{1mm}
\begin{tabular}{ll}
\begin{minipage}{7cm}
\epsfig{file=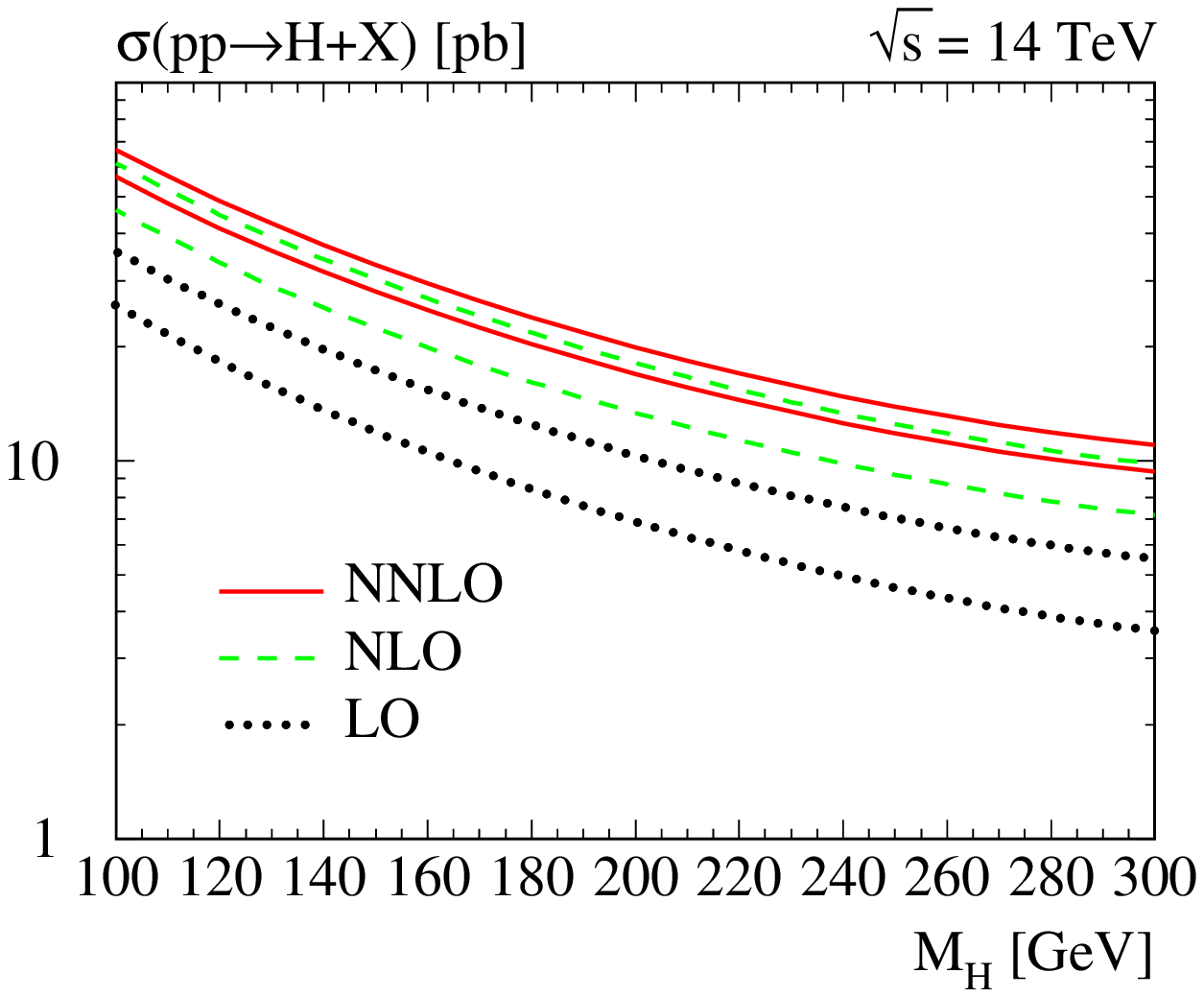,bbllx=110,bblly=265,bburx=465,bbury=560,width=5.6cm} 
\end{minipage}
& 
\hspace*{-10mm}
\begin{minipage}{7cm}
\epsfig{file=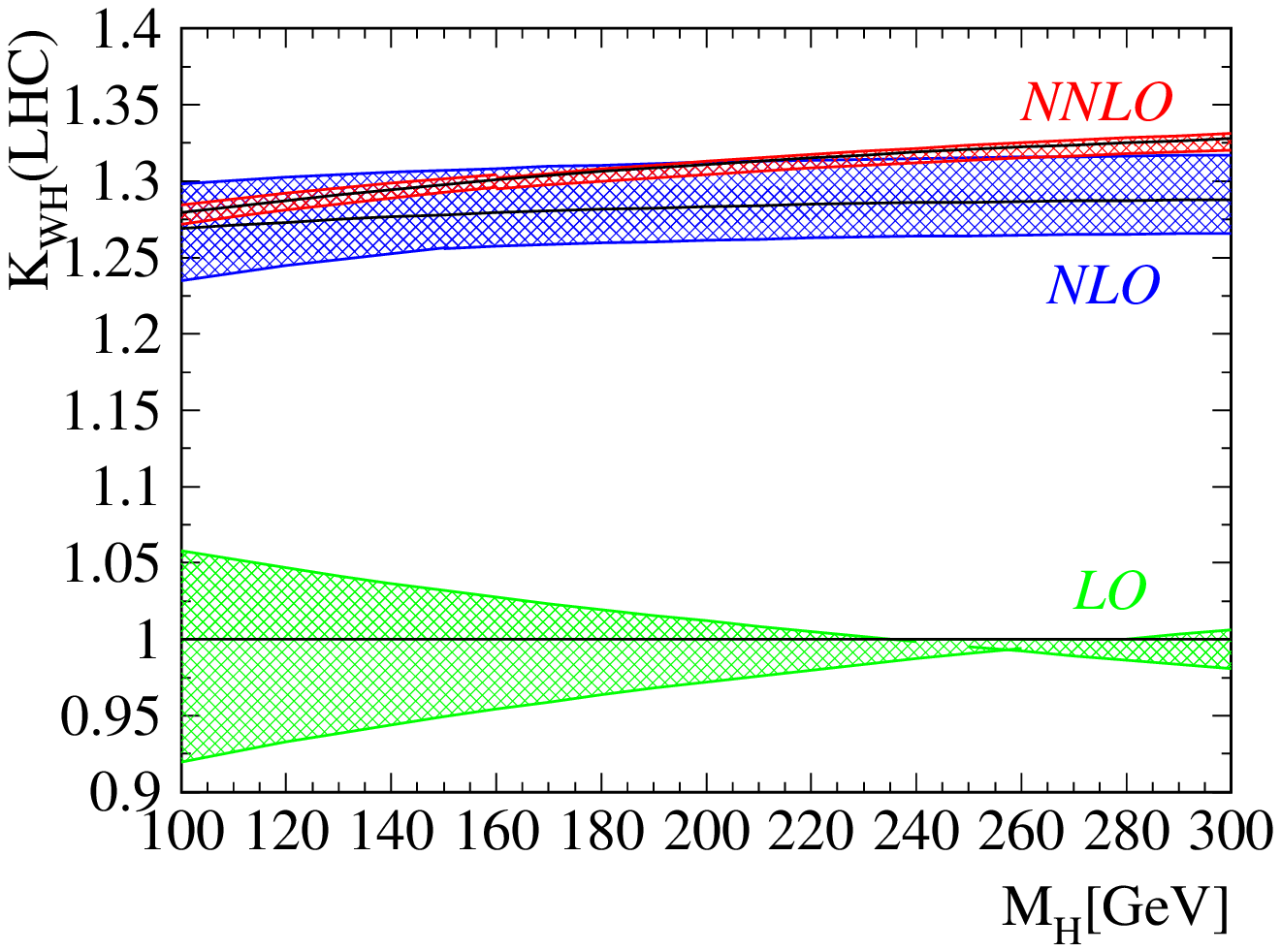, bbllx=110,bblly=265,bburx=465,bbury=560,width=5.6cm} 
\end{minipage}
\end{tabular}
\vspace*{-3mm}    
\end{figure}

{\bf b)} \underline{$q\bar q \to HV$}: The associated production with gauge
bosons, with $H \to b\bar{b}$ and possibly $H \to WW^* \to \ell^+ \nu jj$, is
the most relevant mechanism at the Tevatron \cite{Tevatron}, since the dominant
$gg$ mechanism has too large a QCD background. At the LHC, this process plays
only a marginal role; however, the channels $HW \to \ell \nu \gamma \gamma$ and
eventually $ \ell \nu b\bar b$ could be useful for the measurement of Higgs
couplings.  

The QCD corrections, which at NLO \cite{HV-NLO,HV+VV-NLO}, can be inferred from
Drell--Yan production, have been calculated recently at NNLO \cite{HV-NNLO};
they are of about 30\% in toto (Fig.~7). The ${\cal O} (\alpha)$ electroweak
corrections have been also derived recently \cite{HV-EW} and decrease the rate
by $5$ to 10\%. The remaining scale dependance is very small, making
this process the theoretically cleanest of all Higgs production processes.

{\bf c)} \underline{The $WW/ZZ$ fusion mechanism} has the second largest cross
section at the LHC. The QCD corrections, which can be obtained in the
structure--function approach, are at the level of 10\% and thus small
\cite{VV-NLO,HV+VV-NLO}. The corrections including cuts, and in particular
corrections to the $p_T$ and $\eta$ distributions, have also been calculated
recently and implemented into a parton--level Monte--Carlo program
\cite{VV-MC}. With the specific cuts to the process, the output for the
production cross section is shown in Fig.~8 for a Higgs in the mass range
100--200 GeV.

For several reasons, the interest in this process has grown in recent years: it
has a large enough cross section [a few picobarns for $M_H \lsim 250$ GeV] and
one can use cuts, forward--jet tagging, mini--jet veto for low luminosity as
well as triggering on the central Higgs decay products] \cite{WWfusion0}, which
render the backgrounds comparable to the signal, therefore allowing precision
measurements. In the past, it has been shown that the decay $H \to \tau^+
\tau^-$ and possibly $H \to \gamma \gamma , ZZ^*$ can be detected and could
allow for coupling measurements \cite{Houches,Dieter}. In the last years,
parton--level analyzes have shown that various  other channels can be possibly
detected \cite{WWfusion}:  $H \to WW^*$ for $M_H \sim$ 125--180 GeV, $H \to
\mu^+ \mu^-$ [for  second--generation coupling measurements], $H \to b\bar{b}$
[for the $b\bar{b}H$ Yukawa coupling] and $H \to $ invisible. Recent
experimental  simulations \cite{Karl} have assessed more firmly the potential
of this channel.

\begin{figure}[h] 
\vspace*{-.7cm}
\caption{The $pp\to Hqq$ cross section after cuts as a function of 
$M_H$ at LO (dotted line) and NLO with the tagging jets defined in the $P_T$ 
and $E_T$ methods (left) and the scale variation of the LO and NLO cross 
sections as a function of $M_H$ (right); from Ref.~\cite{VV-MC}.}
\vspace*{2mm}
\centerline{ 
\epsfig{figure=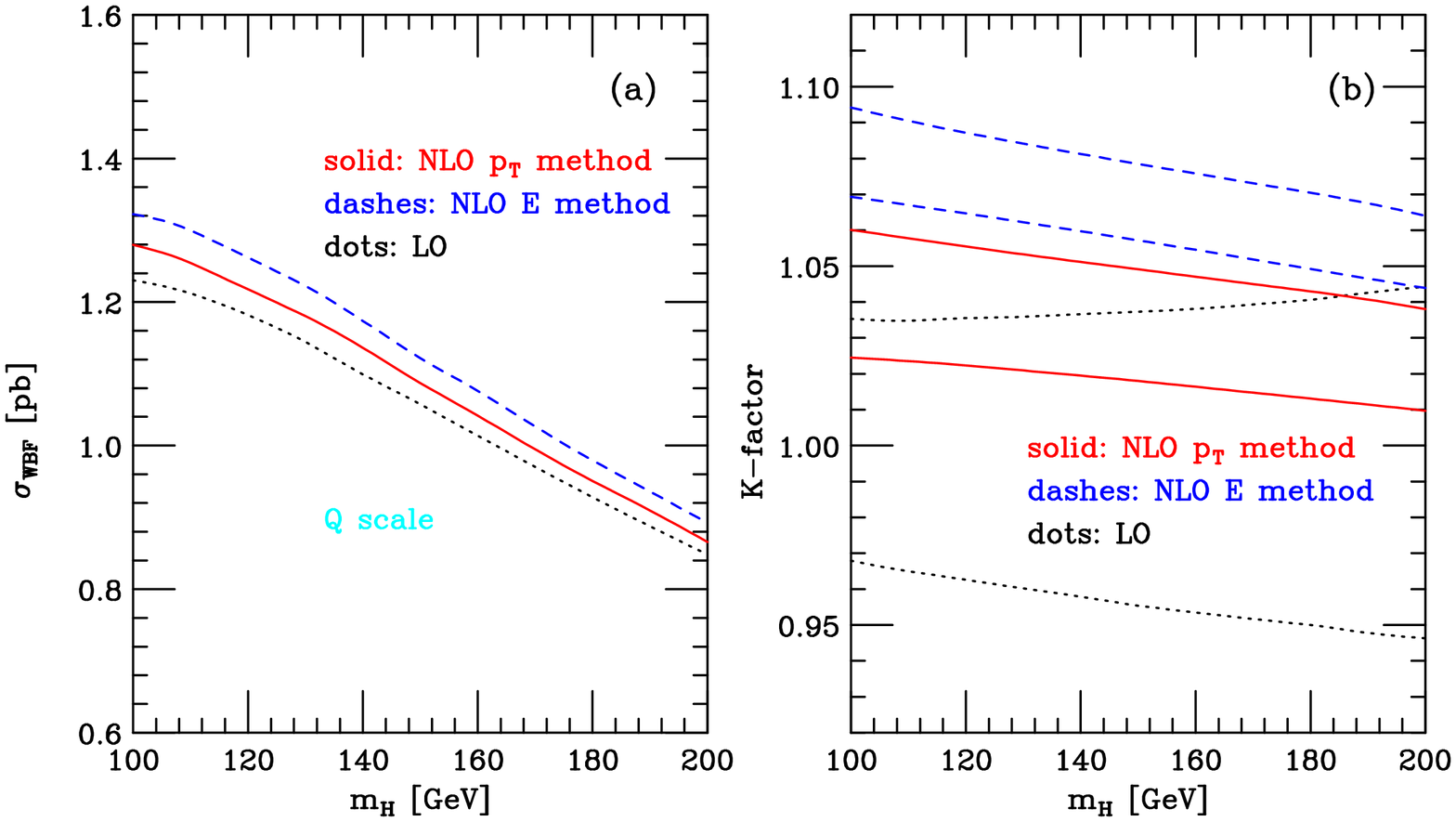,width=0.45 \textwidth, angle=90,clip=} 
} 
\vspace*{-.2cm}
\end{figure}

{\bf d)} Finally, Higgs boson production in association with top quarks, with
$H \to \gamma \gamma$ or $b\bar{b}$, can be observed at the LHC and would allow
for the direct measurement of the top Yukawa coupling. The cross  section is
rather involved at tree--level since it is a three--body process, and the
calculation of the NLO corrections was a real challenge, since one had to deal
with one--loop corrections involving pentagonal diagrams and real corrections
involving four particles in the final state. This challenge was taken up by two
groups [of US ladies and DESY gentlemen], and this calculation was completed
two years ago \cite{ttH}. The $K$--factors turned out to be rather small, $K\sim
1.2$ at the LHC [and $\sim 0.8$ at the Tevatron, an example that  $K$--factors
can also be less than unity]. However, the scale dependence is drastically
reduced from a factor two at LO to the level of 10--20\% at NLO; see Fig.~9 
(left).  Note that the NLO corrections to the $q\bar q /gg \to b\bar b H$ 
process, which is more relevant in the MSSM, have been also recently completed
\cite{bbH}: compared with the NLO rate for the $bg \to bH$ process where the 
initial $b$-quark is treated as a parton \cite{bg-bH}, the calculations agree 
now within the scale uncertainties \cite{bbH-comp} as shown in Fig.~9 (right). 

\nn
\begin{figure}
\vspace*{-1.1cm}
\caption{Left: The production cross sections in  the $t\bar{t}H$ process as a 
function  of the renormalization/factorization scale $\mu$; from 
Dawson et al. in Ref.~\cite{ttH}. Right: the total cross sections for $pp 
\rightarrow b \bar b H+X$ as a function of $M_H$ with one high--$p_T$
$b$ jet identified in the final state and the scales varied by a factor
of two around  $\mu\!=\! \frac{1}{4} M_H$ from \cite{bbH-comp}.}
\vspace*{3mm}
\begin{tabular}{ll}
\begin{minipage}{7.6cm}
\hspace*{-.5cm}
\includegraphics[width=15pc]{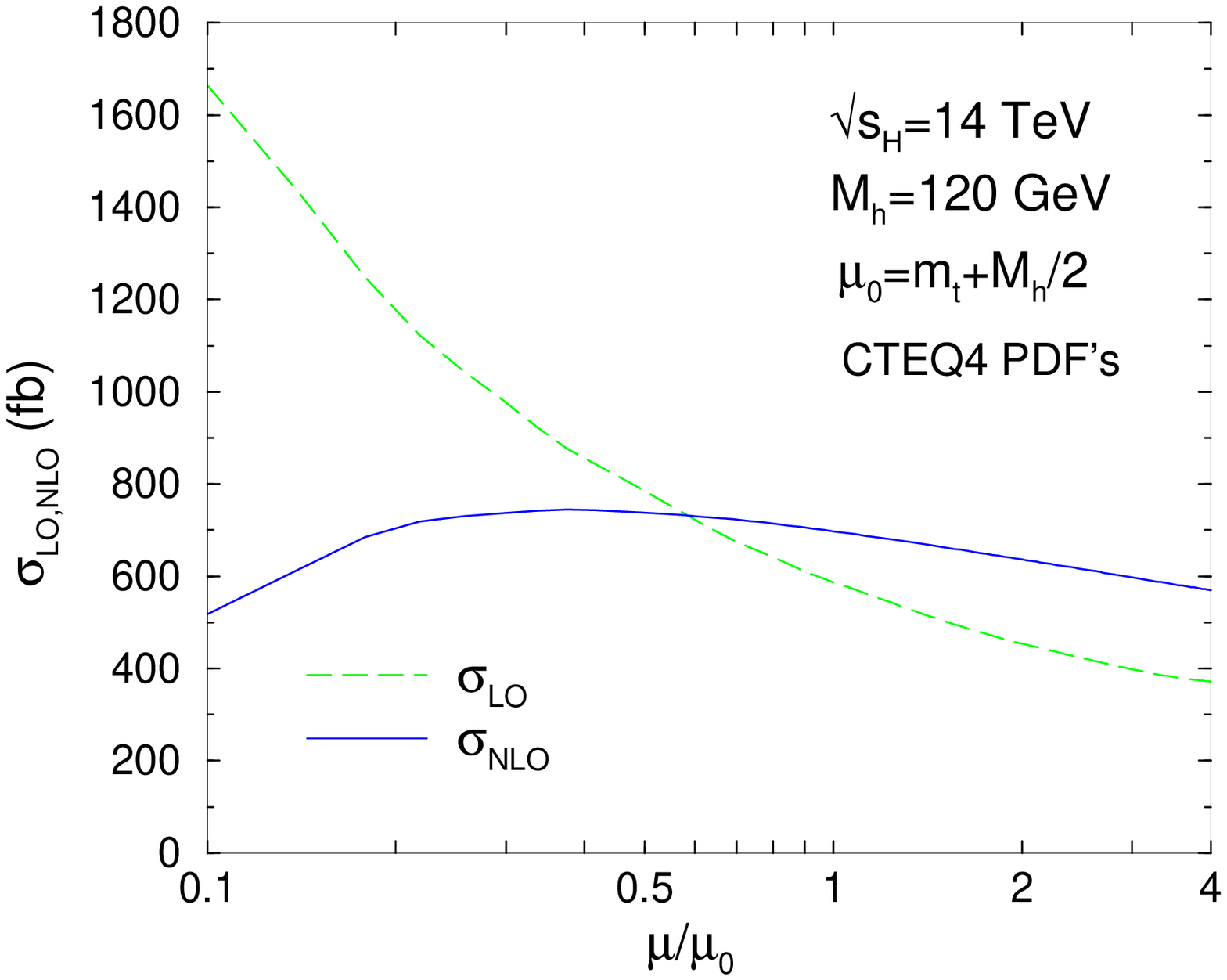}
\end{minipage}
  & \hspace*{-2cm}
\begin{minipage}{7.6cm}
\includegraphics[bb=50 250 580 600,scale=0.35]{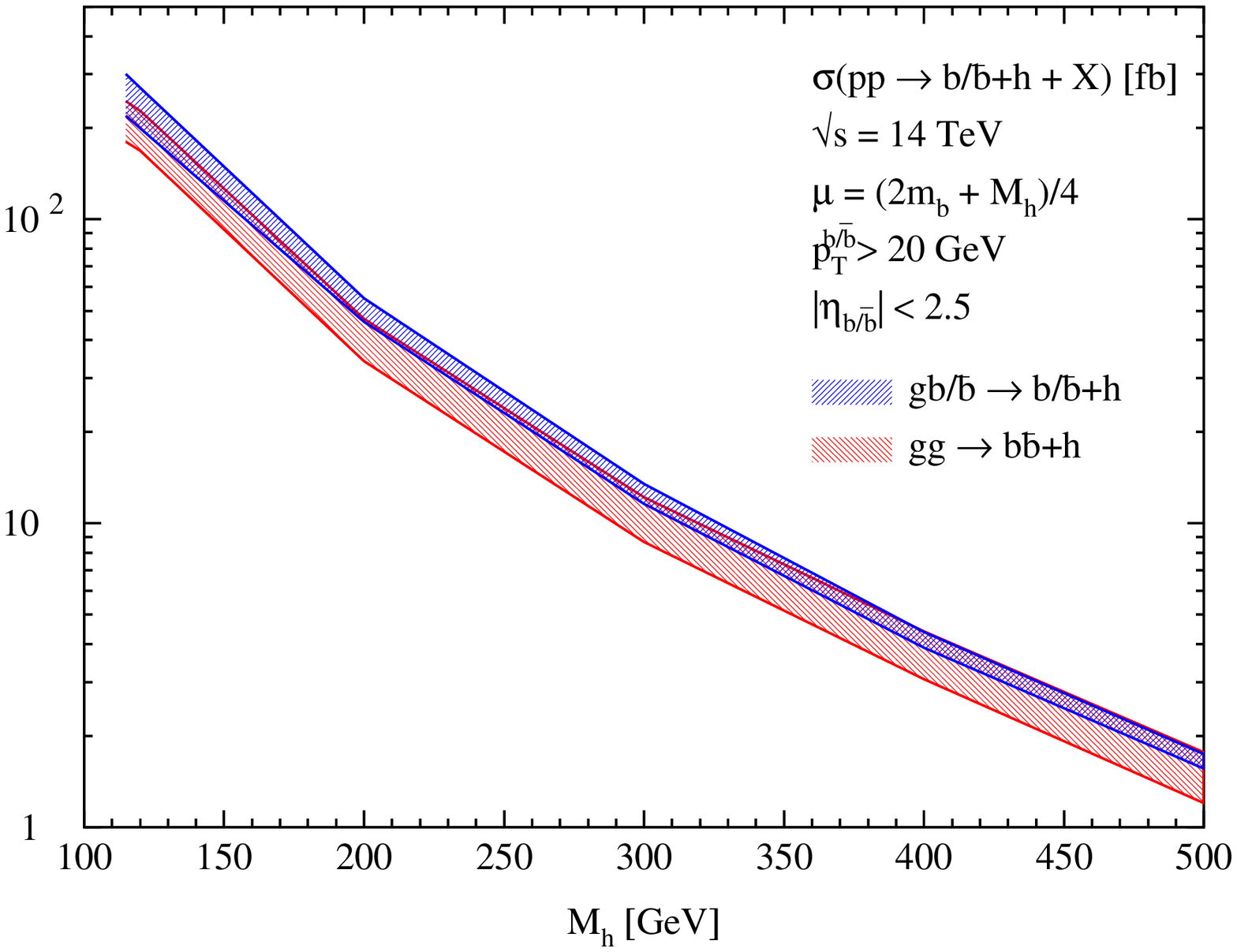}
\end{minipage}
\end{tabular}
\vspace*{-.2cm}
\end{figure}

Note that the PDF uncertainties have also been recently estimated for the four
production processes using the new scheme provided by the CTEQ and MRST
collaborations, as well as by Alekhin \cite{PDF-errors}. At the LHC, the
uncertainties range from 5\% to 15\% depending on the considered process and
the Higgs mass \cite{Samir-PDFs}.  

Let us now turn to the measurements that can be performed at the LHC. 

$\bullet$ The Higgs mass can be measured with a very good accuracy
\cite{LHC,LHCplots}. For $M_H \lsim 400$ GeV, where $\Gamma_H$  is not too 
large, a precision of $\Delta M_H/M_H \sim 0.1$\% can be achieved in $H \to
ZZ^{(*)} \to 4\ell^\pm$.  In the ``low--mass" range, a slight improvement can
be obtained by  considering $H \to \gamma \gamma$. For $M_H \gsim 400$ GeV, the
precision deteriorates because of the smaller rates but it is still at the
level of 1\% up to $M_H\sim 800$ GeV if theoretical errors, such as width
effects, are not taken into account.  

$\bullet$ Using the same process, $H \to ZZ^{(*)} \to 4\ell^\pm$, the total
Higgs width can be measured for $M_H \gsim 200$ GeV, when it is large enough
\cite{LHC,LHCplots}. While the precision is very poor near this mass value [a
factor of two], it improves to reach the level of $\sim 5$\% around $M_H \sim
400$ GeV. Here also, theoretical errors are not included.  

$\bullet$ The Higgs boson spin can be measured by looking at angular
correlations between the fermions in the final states in $H \to VV \to 4f$
\cite{VVHspin}, but the cross sections are rather small and the environment too
difficult; only the measurement of the decay planes of the two $Z$ bosons
decaying into four leptons seems promising.  In vector boson fusion, the
azimuthal distribution of the two tagging jets is different for CP--even and
CP--odd particles and might be used for discrimination \cite{WVB-spin}.
However, if the Higgs boson were a mixture of CP--even and CP--odd states, only
the former component would couple to the gauge bosons. A more decisive test of
the CP numbers should be performed in processes were the Higgs boson couples to
fermions, such as in $pp \to t\bar t H$ with $H\to b\bar b$ as proposed in
Ref.~\cite{ttH-spin}, but this seems too difficult and no experimental analysis
has been attempted yet.  A possibility might be provided by double diffractive
Higgs production with large rapidity gaps between the Higgs and the protons in 
which only scalar Higgs production is selected \cite{diff-spin}.

\begin{figure}[thb]
\vspace*{-1.cm}
\caption{Relative accuracy expected at the LHC with a luminosity of 
200 fb$^{-1}$ for various ratios of Higgs boson partial widths (left) and 
the indirect determination of partial and total widths (right) with some 
theoretical assumptions; from Ref.~\cite{Cpl-measurement}.} 
\vspace*{-.2cm}
\begin{center}
\mbox{
\includegraphics[width=6.cm,angle=90]{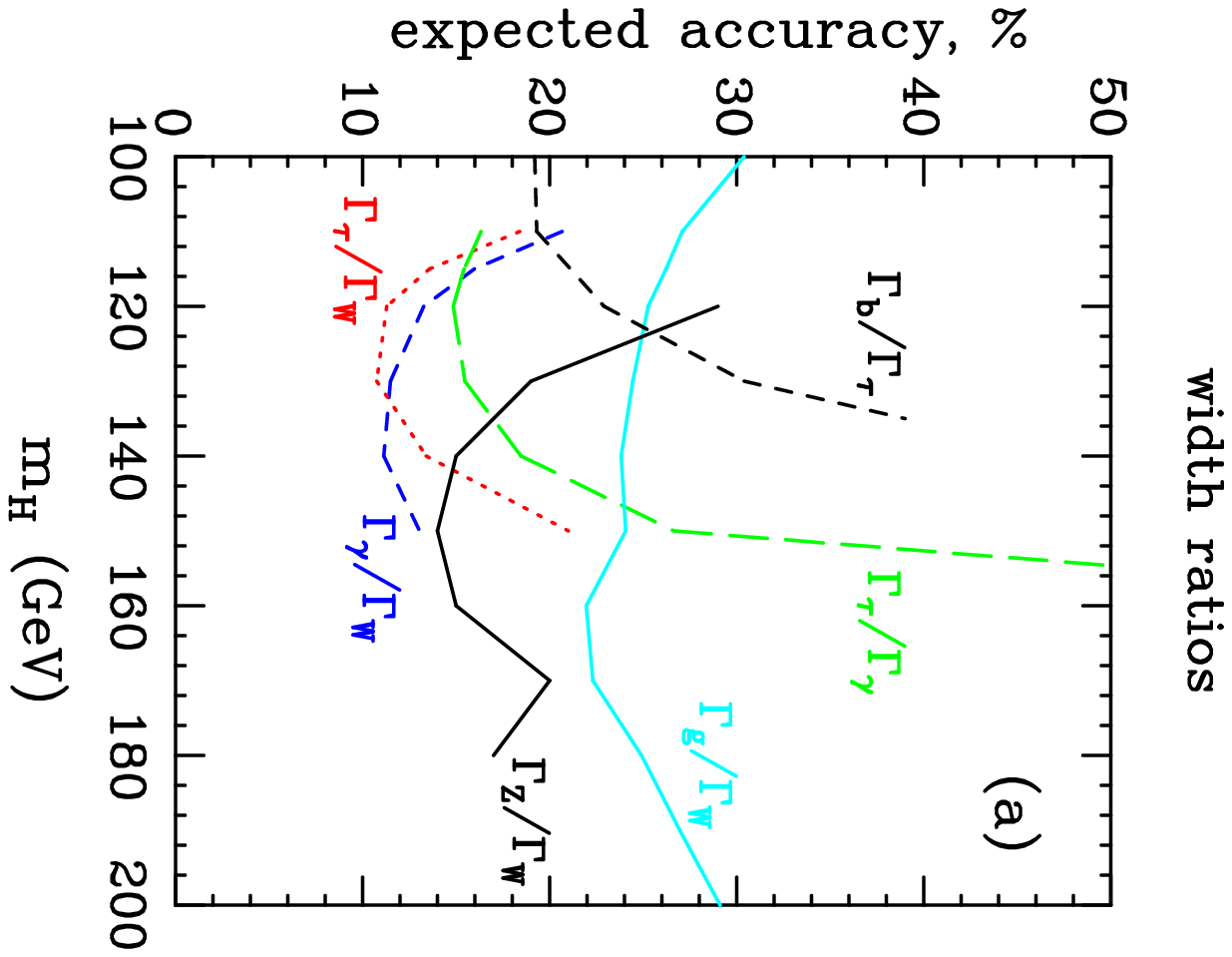} \hspace*{.5cm}
\includegraphics[width=6.cm,angle=90]{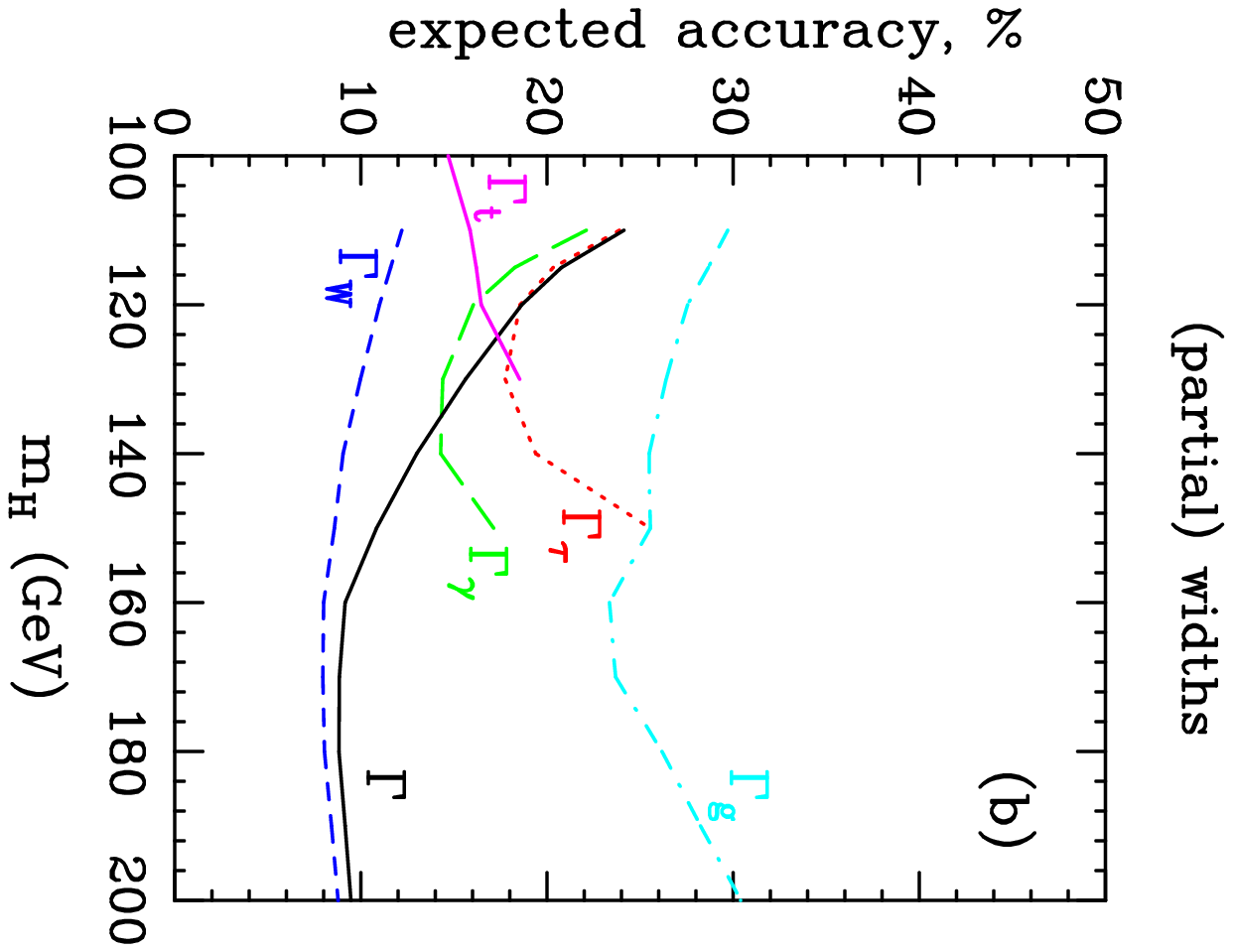} }
\end{center}
\vspace*{-.8cm}
\end{figure}

$\bullet$ The direct measurement of the Higgs couplings to gauge bosons and
fermions is possible, but with a rather poor accuracy as a result of the
limited statistics, the large backgrounds, and the theoretical uncertainties
from the limited precision on the parton densities and the higher--order
radiative corrections.  To reduce some uncertainties,  it is more interesting
to measure ratios of cross sections where the normalizations cancel out. One
can then make, in some cases, a measurement  of ratios of BRs at the level of 
10\% and with some theoretical assumptions, determine the partial and total
widths \cite{Dieter,Karl,Cpl-measurement}.  An example of determination of
cross sections times branching fractions in various channels at the LHC is
shown in Fig.~10.  [Note that experimental analyzes accounting for the
backgrounds and for the detector efficiencies, as well as further theoretical
studies for the signal and backgrounds, need to be performed to confirm these
values.] 

$\bullet$ Finally, the trilinear Higgs self--coupling $\lambda_{HHH}$ is too
difficult to measure at the LHC because of the smallness of the cross sections
\cite{HHHxs} for $gg\to HH$ [and a fortiori the ones for the other channels $VV
\to HH$ and $q \bar q \to HHV$] and the very large backgrounds.  A parton level
analysis has been recently performed in the channel $gg\to HH \to
(W^+W^-)(W^+W^-) \to (jj \ell \nu) (jj \ell \nu)$ and $(jj \ell \nu) (\ell \ell
\nu \nu)$, including all the relevant backgrounds and only at the SLHC with 6
ab$^{-1}$  luminosity that one can hope to determine this coupling but with a
limited accuracy \cite{HHH-SLHC}.

\section{The MSSM Higgs bosons}

In the MSSM, the production processes for the $h,H$ bosons are practically the
same as for the SM Higgs. However, for large $\tb$ values, one has to take the
$b$ quark, whose couplings are strongly enhanced, into account: its loop
contributions in the $gg \to \Phi$ fusion process [and also the extra
contributions from squarks loops, which however decouple for high squark
masses; the SUSY NLO QCD corrections are also available \cite{SUSY-QCD} and are
moderate] and associated production with $b\bar{b}$ pairs, $gg \to b \bar b +
\Phi$ [for which the QCD corrections are available in both the $gg$ and $gb \to
b\Phi, b\bar b \to \Phi$ pictures \cite{bbH,bg-bH,bbH-comp} depending on how
many $b$--quarks are to be tagged, and which are equivalent if the
renormalization and factorization scales are chosen to be small, $\mu  \sim
\frac14 M_\Phi$]. The cross sections for the associated production with
$t\bar{t}$ pairs and $W/Z$ bosons as well as the $WW/ZZ$ fusion processes, are
suppressed for at least one of the particles as a result of the coupling
reduction.  Because of CP invariance, the $A$ boson can be produced only in the
$gg$ fusion and in association with heavy quarks. However, the one--loop
induced processes \cite{gg-AZ,gg-Ag} $gg \to AZ, gg\to Ag$ [which hold for
CP--even Higgsses] and associated production with other Higgs particles,
$pp \to A+h/H/H^+$ \cite{ppHA} are possible but the rates are much smaller in
general.  

The cross sections for the dominant production mechanisms are shown in Fig.~11
as a function of the Higgs masses for $\tb=3$ and $30$ for the same set of
input parameters as Fig.~4. The NLO QCD corrections are included, except 
for the $pp \to \Phi Q \bar Q$ processes where, however, the scales have 
been chosen as to approach the NLO results; the top
mass is fixed to $m_t=178$ GeV and the MRST NLO PDFs have been adopted.  As can
be seen, at high $\tb$, the largest cross sections are by far those of the $gg
\to \Phi_A/A$ and $q\bar q/ gg \to b\bar b+ \Phi_A/A$ processes, where
$\Phi_A=H\, (h)$ in the (anti--)decoupling regimes $M_A > (<) M_h^{\rm max}$:
the other processes involving these two Higgs bosons have cross sections that
are several orders of magnitude smaller. The production cross sections for the
other CP--even Higgs boson, that is $\Phi_H=h\,(H)$ in the (anti--)decoupling
regime when $M_{\Phi_H} \simeq M_h^{\rm max}$, are similar to those of the SM
Higgs boson with the same mass and are substantial in all the channels which
have been displayed. For small values of $\tb$, the $gg$ fusion and $b\bar
b$--Higgs cross sections are not strongly enhanced as before and all production
channels [except for $b\bar b$--Higgs which is only slightly enhanced] have
cross sections that are smaller than in the SM Higgs case, except for $h$ in
the decoupling regime.

For the charged Higgs boson, the dominant channel is the production from top
quark decays, $t \to H^+ b$, for masses not too close to $M_{H^\pm}=m_t\!-\!
m_b$. For higher masses \cite{pp-H+}, the fusion process $gg \to H^\pm tb$
supplemented by $gb \to H^\pm t$ [the two processes have to be properly
combined and the $K$--factor for the $gb$ process has been derived recently
\cite{Kfac-H+} and is moderate $K \sim 1.2$--1.5 if the cross section is
evaluated at scales $\mu \sim \frac12 (m_t+M_{H^\pm})$] are the ones to be
considered. In Fig.~11, shown are the $q\bar q/gg \to H^+ tb$ process which
include the possibility of on--shell top quarks and hence, $pp\to t \bar t$
with $t \to H^+b$. Additional sources \cite{pp-H+-others} of $H^\pm$ states for
masses below $M_{H^\pm} \approx 250$ GeV are provided by pair and associated
production with neutral Higgs bosons in $q\bar q$ annihilation as well as pair
and associated $H^\pm W^\mp$ production in $gg$ and/or $b\bar b$ fusion but the
cross sections are not as large in general.  

\begin{figure}[!h]
\vspace*{-.8cm}
\caption{The cross section for the neutral and charged MSSM Higgs production in
the main channels at the LHC as a function of their respective masses for 
$\tb=3$ and $30$.} 
\begin{center}
\vspace*{-1.6cm}
\hspace*{-3.7cm}
\epsfig{file=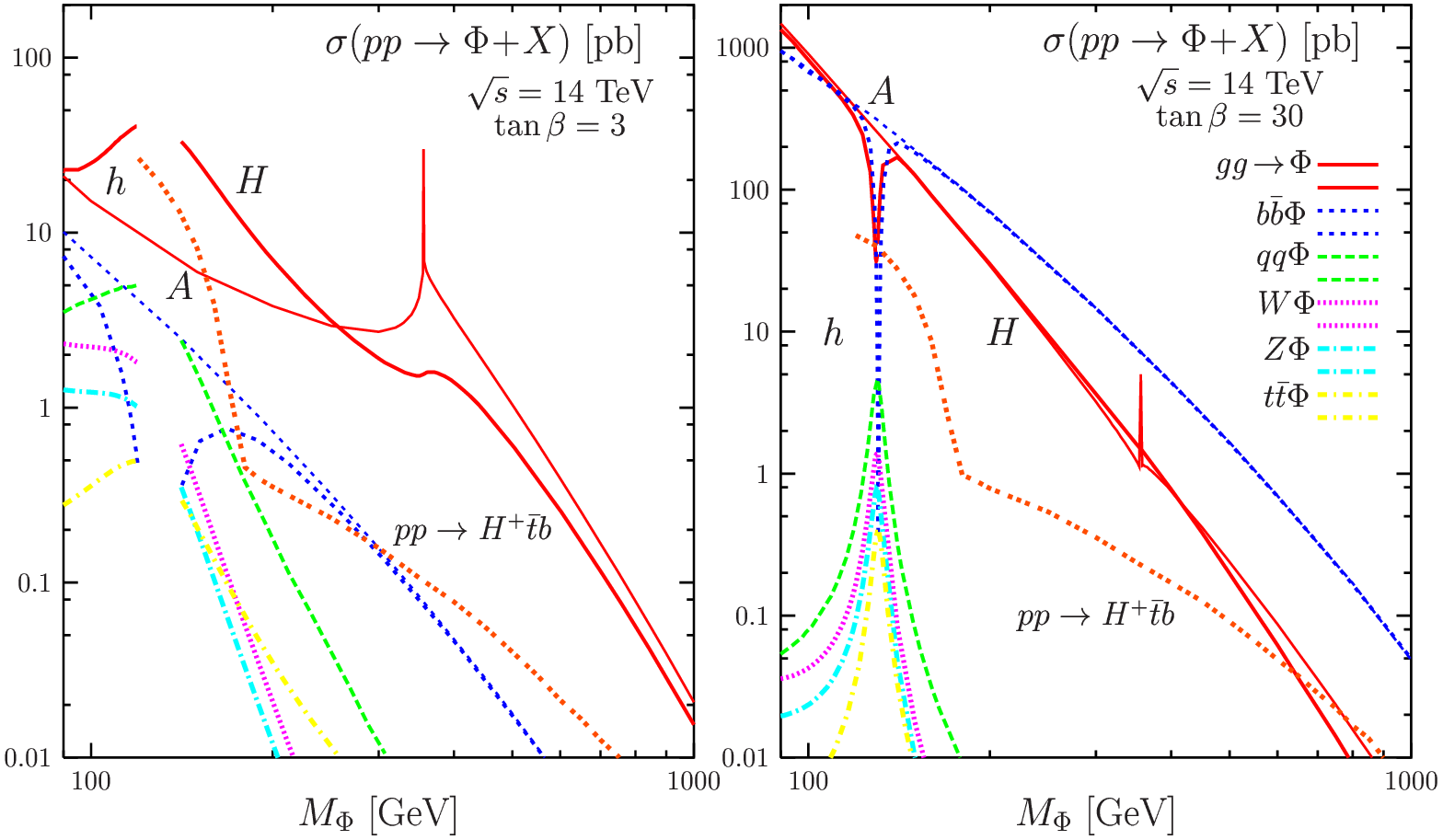,width=10.cm} 
\end{center}
\vspace*{-7.5cm}
\end{figure}

The principal detection signals of the neutral Higgs bosons at the LHC, in the 
various regimes of the MSSM, are as follows 
\cite{LHC,Houches,Houches03,MSSM-A,MSSM-C,intense}:  

In the \underline{decoupling regime}, i.e. when $M_h \simeq M_h^{\rm max}$, the
lighter $h$ boson is SM--like and has a mass smaller than $\approx 140 $ GeV. 
It can be detected in the $h \to \gamma \gamma$ decays [possibly supplemented
with a lepton in associated $Wh$ and $t\bar t h$ production], and eventually in
$h\to ZZ^*, WW^*$ decays in the upper mass range, and if the vector boson
fusion processes are used, also in the decays $h \to \tau^+ \tau^-$ and
eventually $h \to W W^*$ in the higher mass range $M_{h} \gsim 130$ GeV; see
Fig.~12.  For relatively large values of $\tb$ $(\tb \gsim 10)$, the heavier
CP--even $H$ boson which has enhanced couplings to down--type fermions, as well
as the pseudoscalar Higgs particle, can be observed in the process $pp \to
b\bar b + H/A$ where at least one $b$--jet is tagged and with the Higgs boson
decaying into $\tau^+ \tau^-$, and eventually, $\mu^+ \mu^-$ pairs in the low
mass range. With a luminosity of 100 fb$^{-1}$ [and is some cases lower] a
large part of the $\tb$--$M_A$ space can be covered; Fig.\,12.  

In the \underline{antidecoupling regime}, i.e. when $M_A < M_h^{\rm max}$ and
at high $\tb$ ($\gsim 10$), it is the heavier $H$ boson which will be
SM--like and can be detected as above, while the $h$ boson will behave like the
pseudoscalar Higgs particle and can be observed in $pp \to b\bar b+ h$ with $h
\to \tau^+ \tau^-, \mu^+ \mu^-$ provided its mass is not too close to $M_Z$ not
to be swamped by the background for $Z$ production. The part of the $\tb$--$M_A$
space which can be covered is also shown in the left--hand side of Fig.~12.  

In the \underline{intermediate coupling regime}, that is for not too large
$M_A$ values and moderate $\tb \lsim 5$, the interesting decays $H \ra hh$, $A
\ra hZ$ and even $H/A \ra t\bar{t}$ [as well as the decays $H^\pm \to Wh$]
still have sizable branching fractions and can be searched for; Fig.~13 (left). 
In particular, the $gg \to H \to hh \to b\bar b \gamma \gamma$ process [the
$4b$ channel is more difficult] is observable for $\tb \lsim 3$ and $M_A \lsim
300$ GeV, and would allow to measure the trilinear $Hhh$ coupling.
These regions of parameter space have to be reconsidered in the light of the
new Tevatron value for the top quark mass and the recent analyzes which have
re-opened the small $\tb$ window.  

\begin{figure}[!h]
\vspace*{-.9cm}
\caption{The areas in the $(M_A, \tb)$ parameter space where the lighter (left)
and heavier (right) MSSM neutral Higgs bosons can be discovered at the LHC with
an integrated luminosity of 30 fb$^{-1}$ in the standard production channels; 
from \cite{MSSM-C}.}
\vspace*{-.4cm}
\begin{center}
\mbox{
\includegraphics[width=6.cm,height=5.cm]{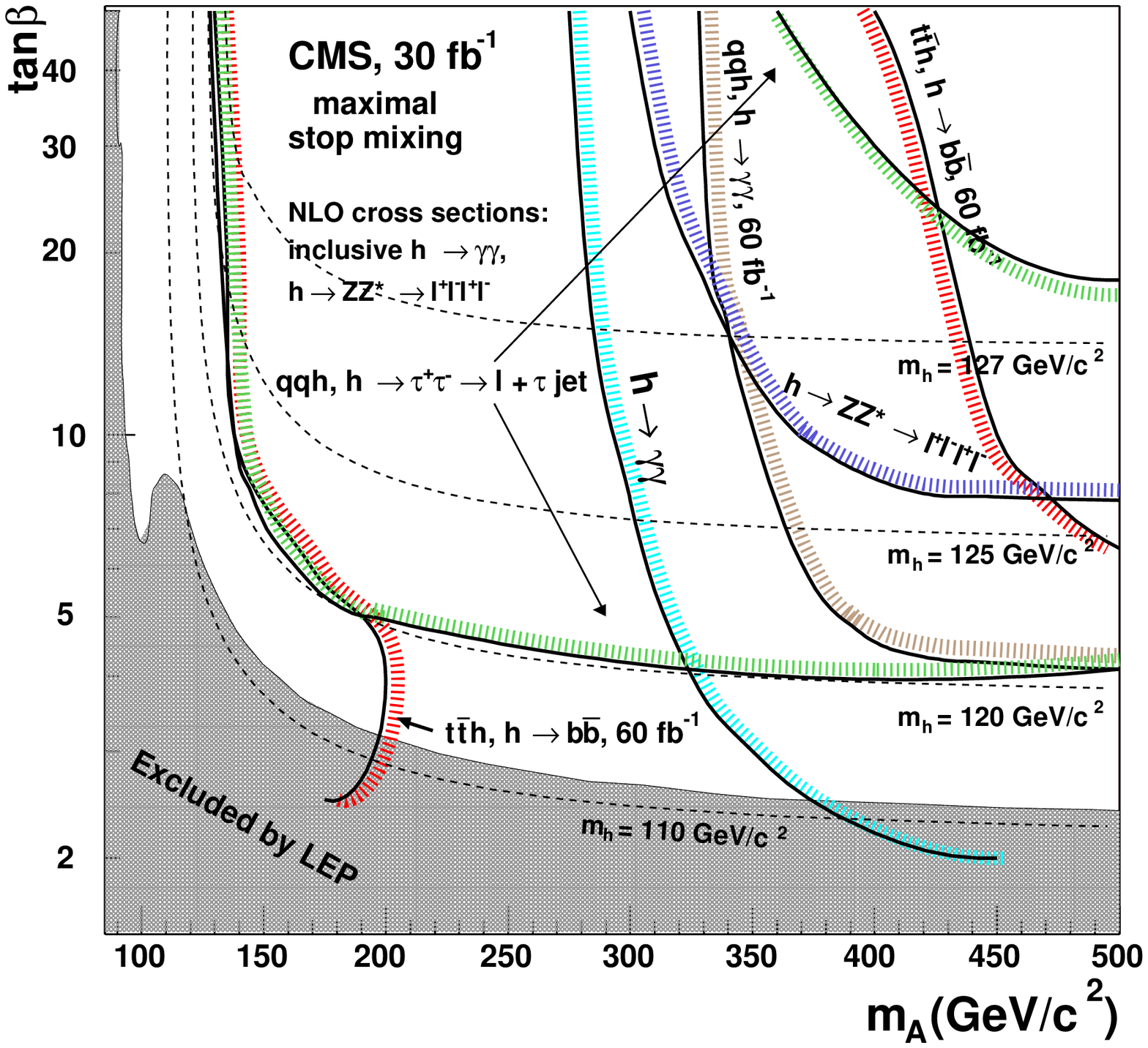} 
\includegraphics[width=6.cm,height=5.cm]{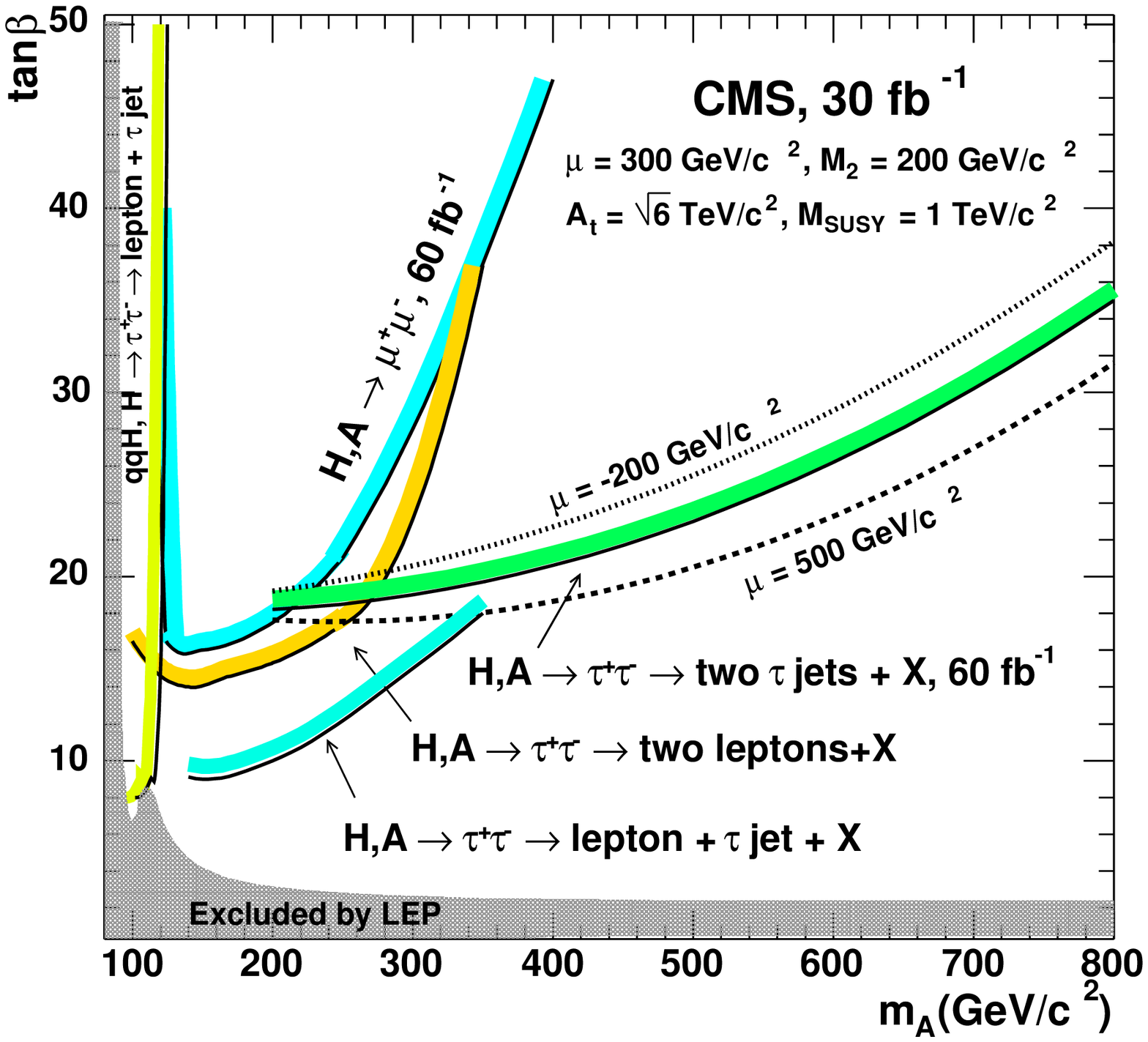} }
\end{center}
\vspace*{-.9cm}
\end{figure}

\begin{figure}[!h] 
\vspace*{-5mm}
\caption{Left: the regions in the $M_A$--$\tb$ parameter space where the
channel $gg \to H \to hh \to b\bar b \gamma \gamma$, $gg \to A \to hZ \to
b\bar b \ell^+ \ell^-$ and $gg \to H/A \to t\bar t \to \ell \nu jj 
b\bar b $ can be detected at the LHC; from Ref.~\cite{MSSM-A}. Right: the
$\mu^+ \mu^-$ pair invariant mass distributions for the three Higgs signal 
peaks with $M_A=125$ GeV and $\tb=30$ [leading to $M_h \sim 124$ GeV and $M_H 
\sim 134$ GeV] and backgrounds after detector resolution smearing; from 
Ref.~\cite{intense}.}
\vspace*{-8mm}
\begin{center}
\begin{tabular}{cc}
\begin{minipage}{7cm}
\hspace*{5mm}
\mbox{
\includegraphics[width=3cm,height=2.2cm]{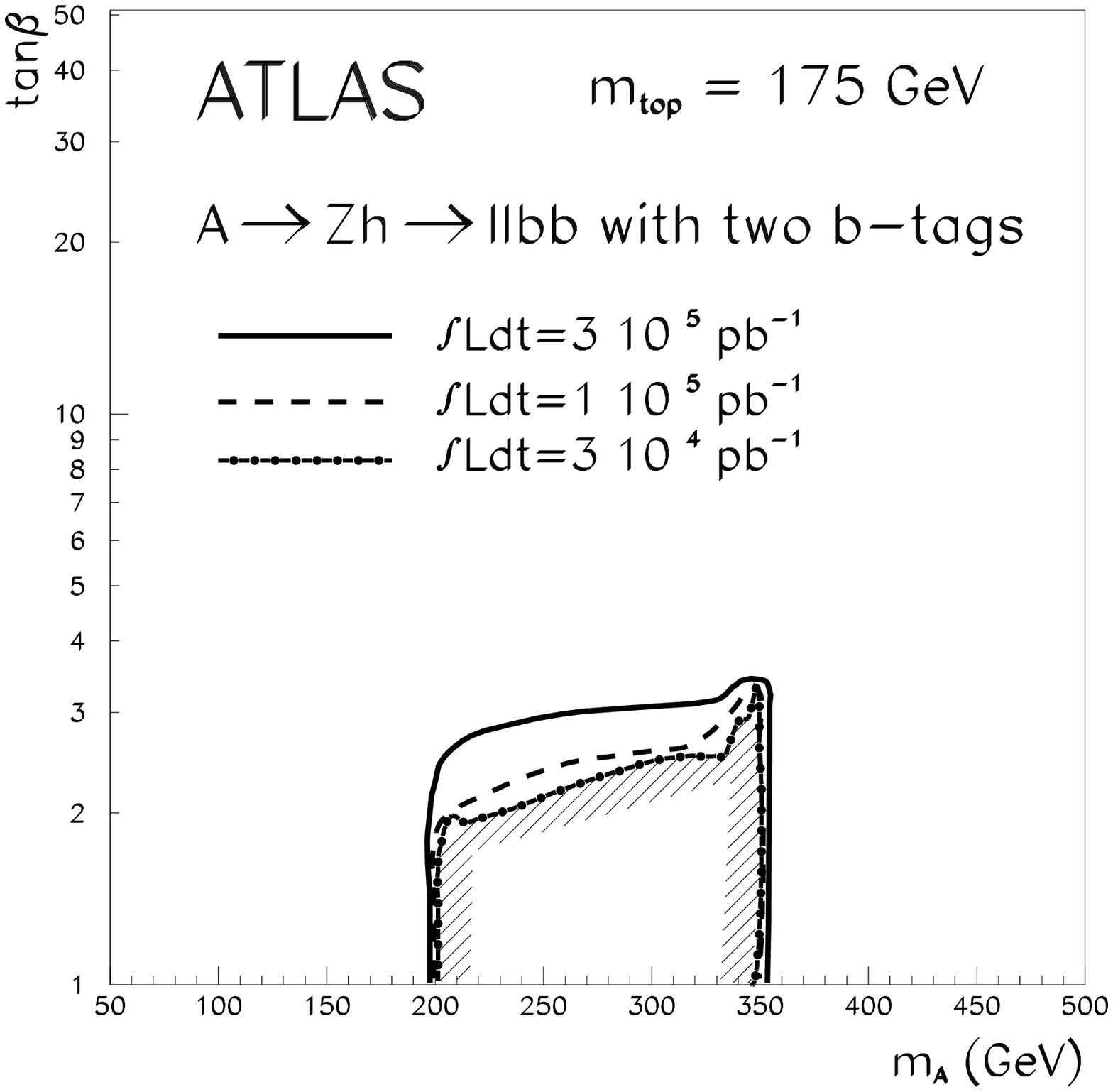} \hspace{-2mm}
\includegraphics[width=3cm,height=2.2cm]{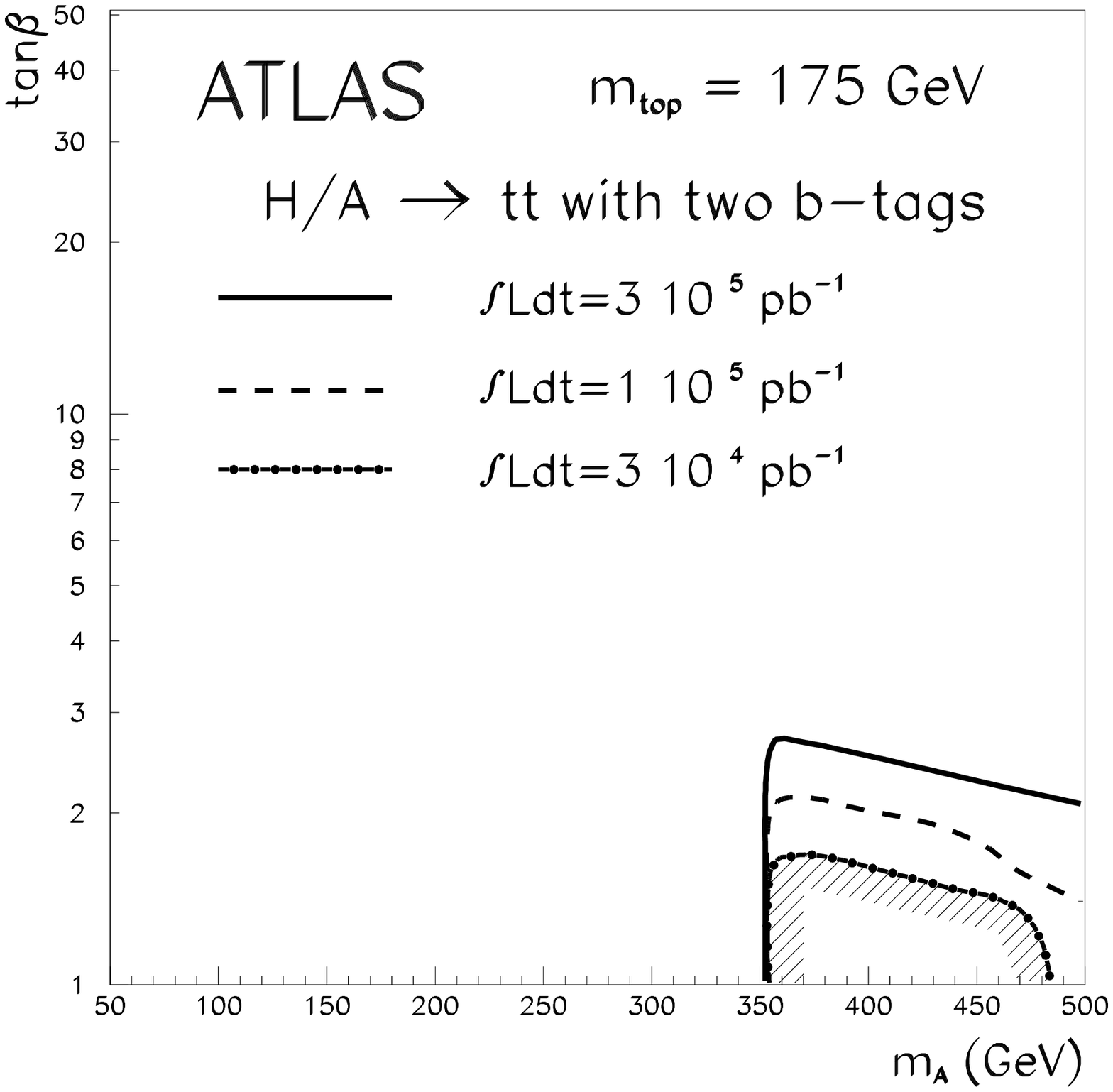}}\\[2mm]
\hspace*{18mm}
\mbox{
\includegraphics[width=4cm,height=2.2cm]{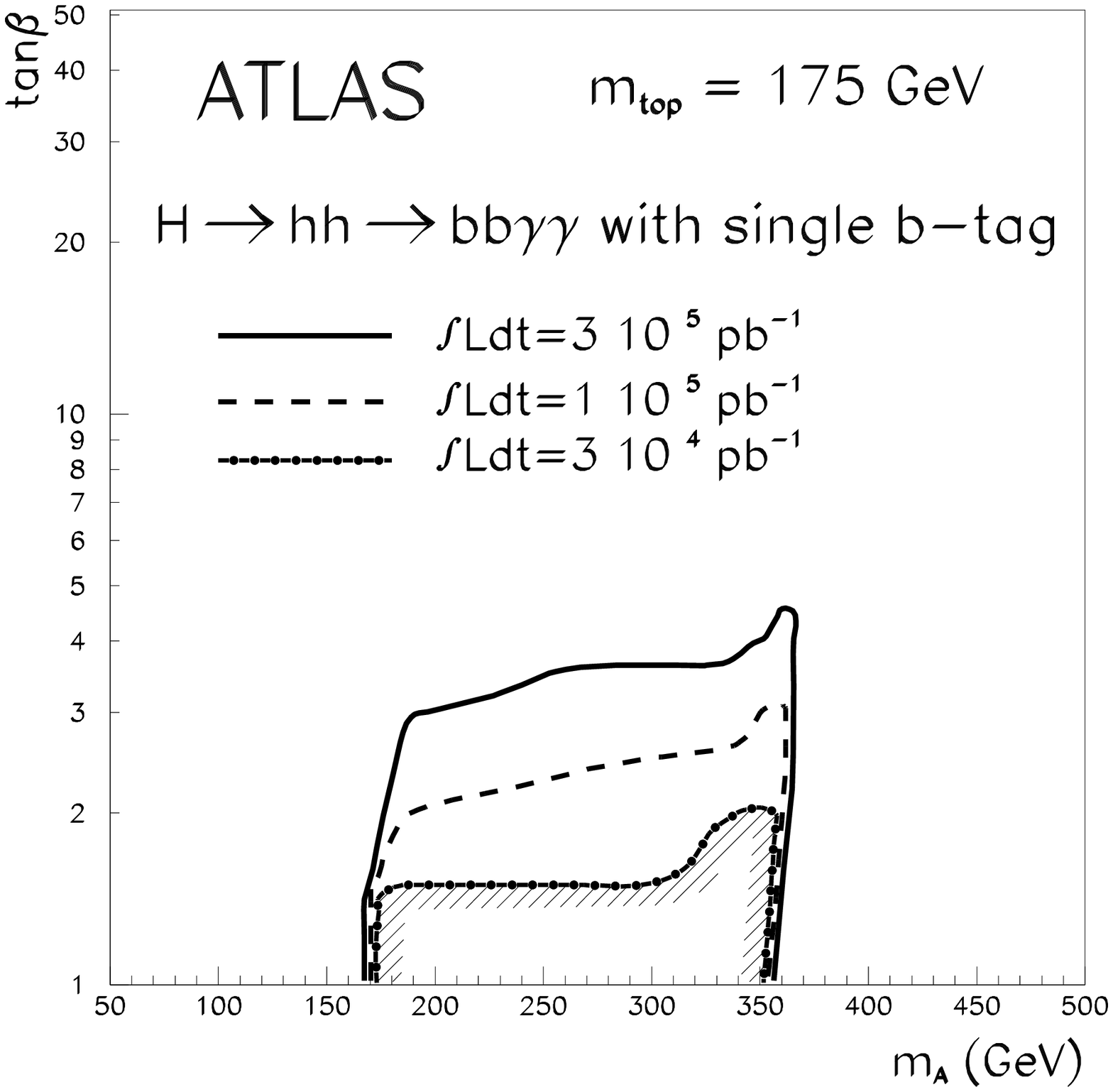} \hspace{-2mm}
}
\end{minipage}
&
\begin{minipage}{7cm}
\hspace*{-5mm}
\includegraphics[width=5.3cm,height=5.3cm]{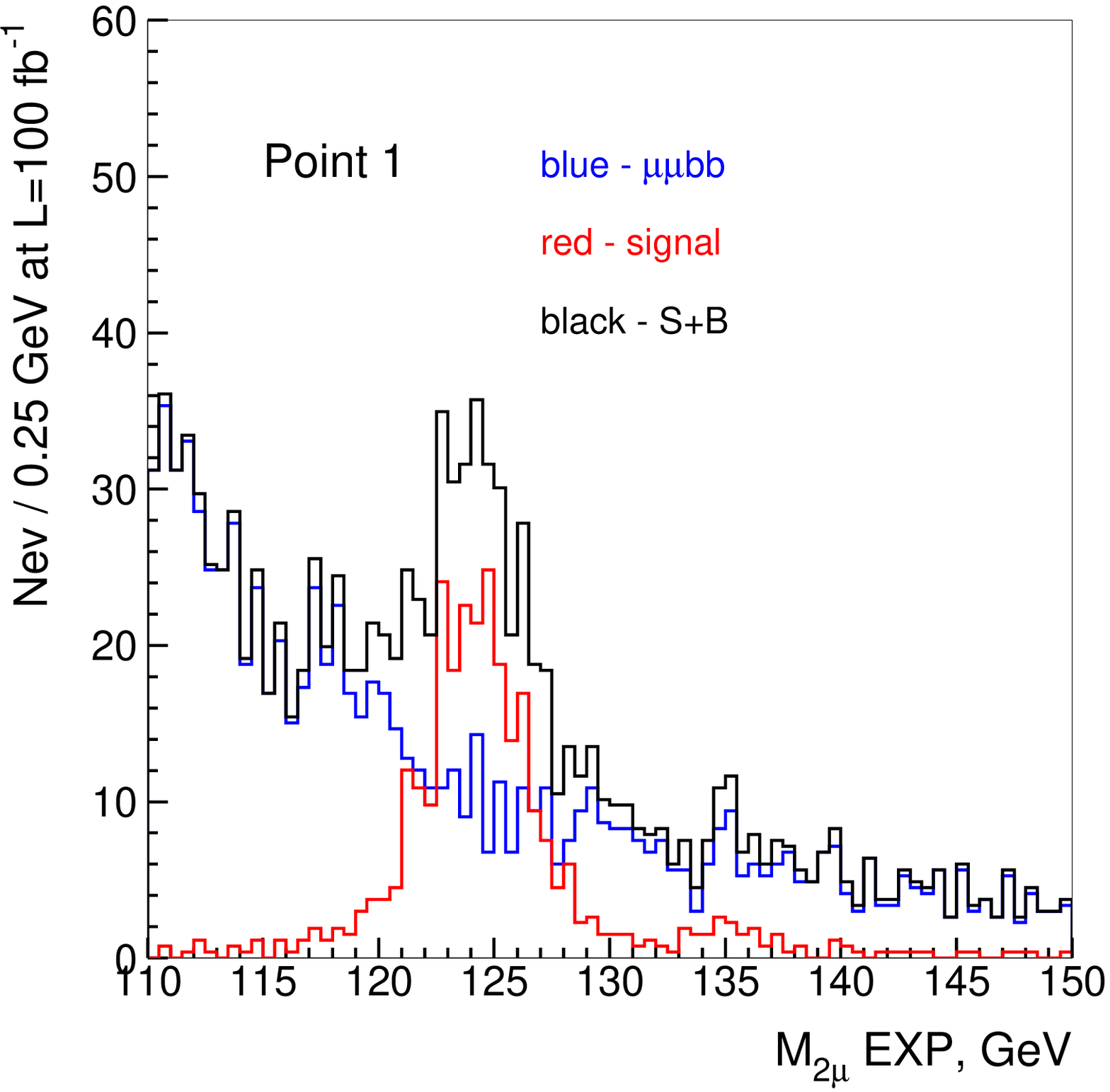} 
\end{minipage}
\end{tabular}
\end{center}
\vspace*{-11mm}
\end{figure}

In the \underline{intense--coupling regime}, that is for $M_A \sim M_h^{\rm
max}$ and $\tb \gg1$, the three neutral Higgs bosons $\Phi=h,H,A$ have
comparable masses and couple strongly to isospin $-\frac{1}{2}$ fermions
leading to dominant decays into $b\bar b$ and $\tau\tau$ and large total decay
widths \cite{intense,intense0}. The three Higgs bosons can only be produced in
the channels $gg \to \Phi$ and $gg/q\bar q \to b\bar b + \Phi$ with $\Phi \to
b\bar b, \tau^+\tau^-$ as the interesting $\gamma \gamma, ZZ^*$ and $WW^*$
decays of the CP--even Higgsses are suppressed. Because of background and
resolution problems, it is very difficult to resolve between the three
particles. A solution advocated in \cite{intense} (see also \cite{ggmumu}),
would be the search in the channel $gg/q\bar q \to b\bar b + \Phi$ with the
subsequent decay $\Phi \to \mu^+ \mu^-$ which has a small BR,
$\sim 3 \times 10^{-4}$, but for which the better muon resolution, 
$\sim 1\%$, would allow to disentangle between at least two Higgs particles. 
[The backgrounds are much larger for the $gg \to \Phi \to \mu^+ \mu^-$ signal].
The simultaneous discovery of the three Higgs particles is very difficult and
in many cases impossible, as exemplified in Fig.~13 (right) where one observes
only one single peak corresponding to $h$ and $A$ production.  

Finally, as mentioned previously, light $H^\pm$ particles with masses below
$M_{H^\pm} \sim m_t$ can be observed in the decays $t \ra H^+b$ with $H^-\ra
\tau \nu_\tau$, and heavier ones can be probed for large enough $\tb$, by
considering the properly combined $gb \to t H^-$ and $gg \ra t \bar{b} H^-$
processes using the decay $H^-\ra \tau \nu_\tau$ and taking advantage of the
$\tau$ polarization to suppress the backgrounds, and eventually the decay $H^-
\to \bar{t}b$ \cite{Siannah} which however, seems more problematic than thought
as recently pointed out \cite{Steven}. See Ref.~\cite{DP} for more detailed
discussions on $H^\pm$ production.  

The whole discussion made previously assumes that Higgs decays into SUSY
particles are kinematically inaccessible. This seems to be unlikely since at
least some charginos  and neutralinos should be not too heavy \cite{J-F.G} and
in this \underline{SUSY regime}, $\Phi \to \chi \chi$ decays are possible
\cite{SUSYdecays}.  Preliminary analyzes show that  decays  $H/A \to \chi_2^0
\chi_2^0 \to 4  \ell^\pm X$ and $H^\pm \to \chi_2^0 \chi_1^\pm \to 3 \ell^\pm X
$ can be detected in some cases; see the l.h.s of Fig.~14. It is also possible
that the lighter $h$ decays invisibly into the lightest neutralinos [or
sneutrinos]; if this scenario is realized, the discovery of these Higgs
particles will be challenging but possible \cite{Houches03}.  Light SUSY
particles can also alter the loop--induced production and decay rates. For
instance, light top squarks can couple strongly to the $h$ boson, leading to a
possibly drastic suppression of the product $\sigma( gg \to h)\times {\rm BR}
(h \to \gamma \gamma)$ compared to the SM \cite{SUSYloops}. In this case,
associated $\tilde t_1 \tilde t_1 h$ production might be accessible with
reasonable rates \cite{SUSYdirect}.

MSSM Higgs boson detection from the cascade decays of strongly interacting
sparticles, which have large production rates at the LHC, is also possible.  In
particular, the lighter $h$ boson and the heavier $A,H$ and $H^\pm$ particles
with $M_\Phi \lsim 200$ GeV, can be produced from the decays of squarks and
gluinos into the heavier charginos/neutralinos, which then decay into the
lighter ones and Higgs bosons. This can occur either in ``little cascades",
$\chi_2^0, \chi_1^\pm \to \chi_1^0 + \Phi$, or in ``big cascades"
$\chi_{3,4}^0, \chi_2^\pm \to \chi_{1,2}^0, \chi_1^\pm + \Phi$.  Recent studies
\cite{cascade0,cascade} show  that these processes can be complementary to
the direct production ones  in some areas of the MSSM parameter space [in
particular one can probe the region $M_A \sim 150$ GeV and $\tb \sim 5$, where
only the $h$ boson can be observed in  standard searches]; see Fig.~14 (right).

\begin{figure}[!h]
\vspace*{-1.cm}
\caption{Areas in the $(M_A, \tb)$ parameter space where the MSSM Higgs bosons 
can be discovered at the LHC with 100 fb$^{-1}$ data in the $A/H \ra \chi^0_2 
\chi^0_2\ra 4\ell^\pm+X$ decays (left) and in cascades of SUSY particles (right)
for a given set of the MSSM parameters.} 
\vspace*{-.9cm}
\begin{center}
\mbox{
\epsfig{file=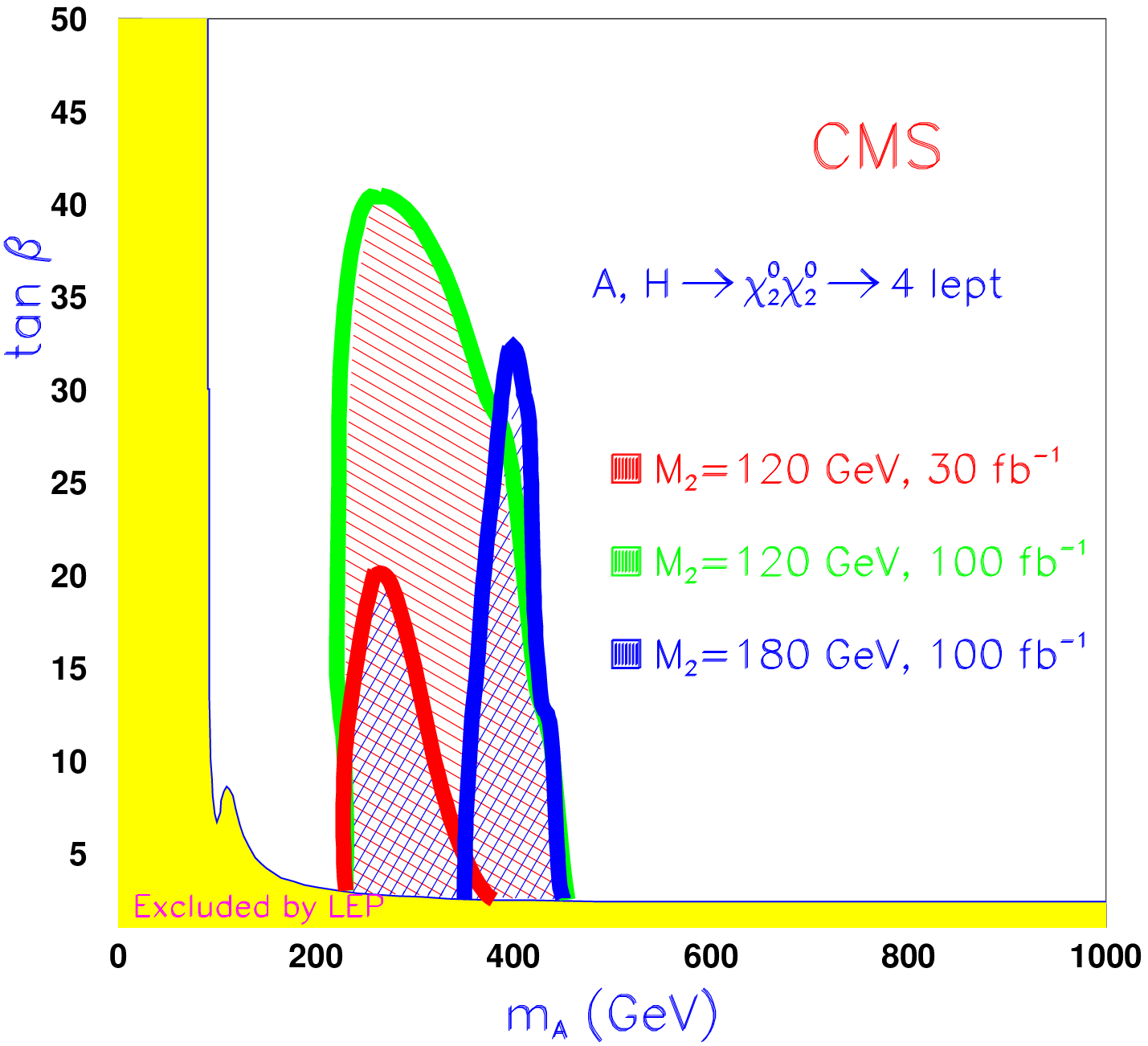,width=6.2cm}
\epsfig{file=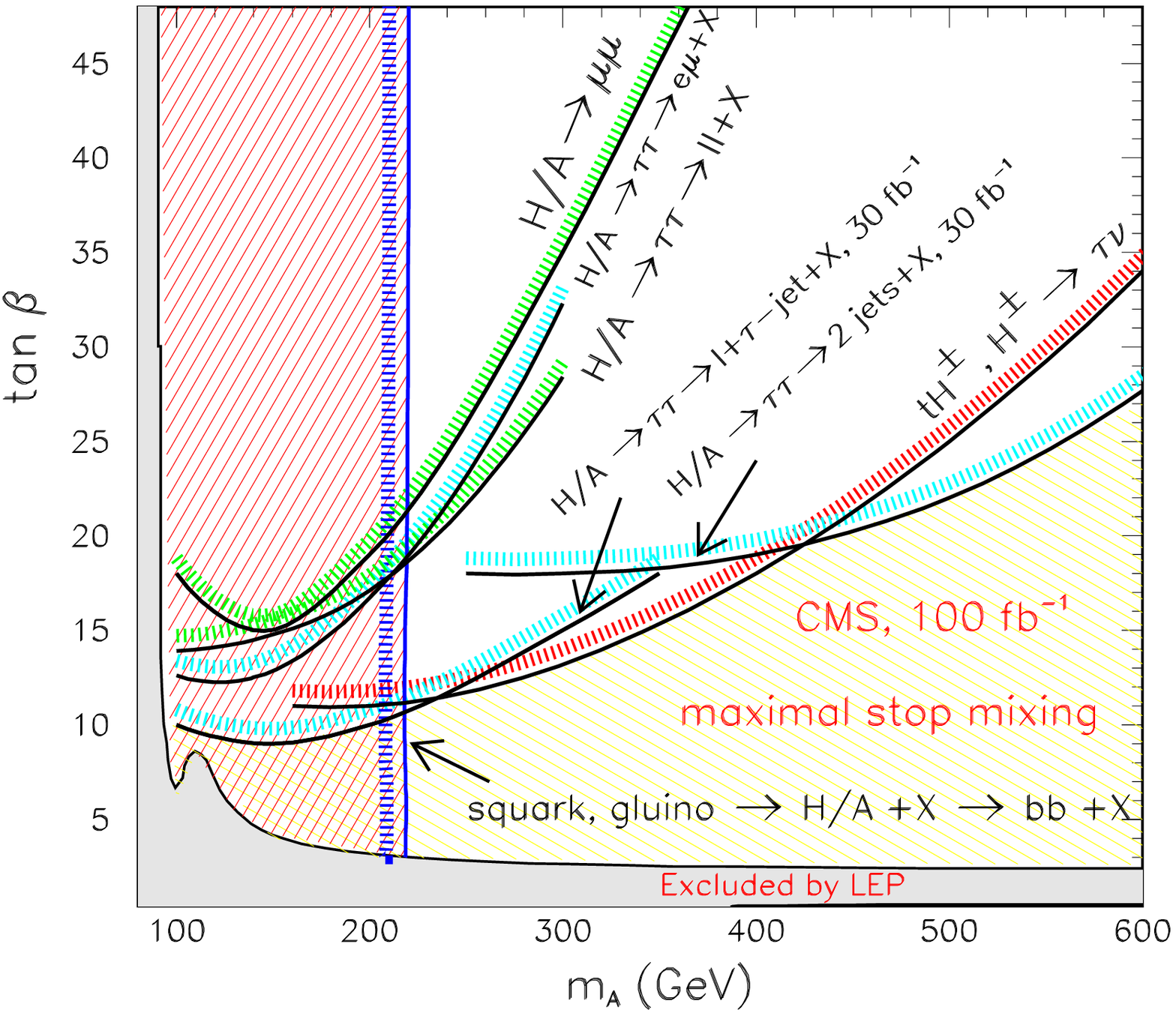,width=6.2cm} } 
\end{center}
\vspace*{-.7cm}
\end{figure}

Finally, there is the \underline{extended regime} in which some basic
assumptions of the MSSM are relaxed. In the presence of CP--violation in the
SUSY sector, the new phases will enter the MSSM Higgs sector through the large
radiative corrections. The masses and the couplings of the neutral and charged
Higgs particles will be altered [in particular, the three neutral Higgs
bosons will not have definite CP quantum numbers and will mix with each
other] and their production cross sections and decay branching ratios will be
affected; for a review, see e.g.~Ref.~\cite{cpxbenchmark}. The impact of
CP--violation on the LEP2 bounds on the Higgs masses has been recently
evaluated \cite{cpxopal} and the discovery potential of the LHC has been studied
\cite{schumacher} for several benchmarks \cite{cpxbenchmark}.  An
ATLAS simulation with 300 fb$^{-1}$ data has shown that the lighter neutral
Higgs boson can escape observation in a small region of the parameter space
with low $M_A$ and $\tb$ values, while the heavier $H,A$ and $H^\pm$ bosons 
can be accessed in smaller areas than in the usual MSSM; Fig.~15 (left).  

Another extension of the MSSM which started to be studied intensively is the
NMSSM where a singlet superfield is introduced, leading to an additional
CP--even and CP--odd Higgs particles \cite{nmssm}. The upper bound on the mass
of the lighter CP--even Higgs particle of the model exceed that of the MSSM $h$
boson and the negative searches at LEP2 lead to looser constraints on the
masses. A new code, {\tt NMHDECAY} \cite{Cyril}, which determines the Higgs
spectrum in the model has appeared recently. Several new decay channels take
place which complicate the searches at the LHC \cite{nmssm1}. For instance, the
lighter CP--even $h_1$ boson could decay into two pseudoscalar $a_1$ bosons
which have masses of order of a few ten GeV, that are not excluded at LEP2.  If
such a scenario occurs, it would be extremely difficult to access to these
particles at the LHC as shown in an ATLAS analysis summarized in Fig.~15
(right), which compares the signal cross sections in the fusion process $pp \to
h_1 qq$ with the various backgrounds \cite{Stephanie}. More detailed studies
are thus needed in this context.  

\begin{figure}[!h]
\vspace*{-1.cm}
\caption{Left: the overall discovery potential for Higgs bosons in ATLAS in a
CP--violating scenario after collecting 300 fb$^{-1}$ of data, with the white
region indicating the area where no Higgs boson can be found; from
\cite{schumacher}. Right: the signal and backgrounds as a function of the
invariant  mass $M_{bb \tau \tau}$ in GeV for the production of a Higgs boson
in the NMSSM in the reaction $pp \to qq+ h_1$ with $h_1 \to a_1 a_1 \to b\bar b
\tau^+ \tau^-$; the signal $\times 500$ (blue),  $t\bar t$ (purple), $\gamma ^*
\to e^+e^-, \, \mu \mu$ (green), $Z \to \tau^+ \tau^-$ (red) and 
total background (black); from \cite{Stephanie}.}
\vspace*{-1.1cm}
\begin{center}
\begin{tabular}{ll}
\begin{minipage}{6.5cm}
\epsfig{file=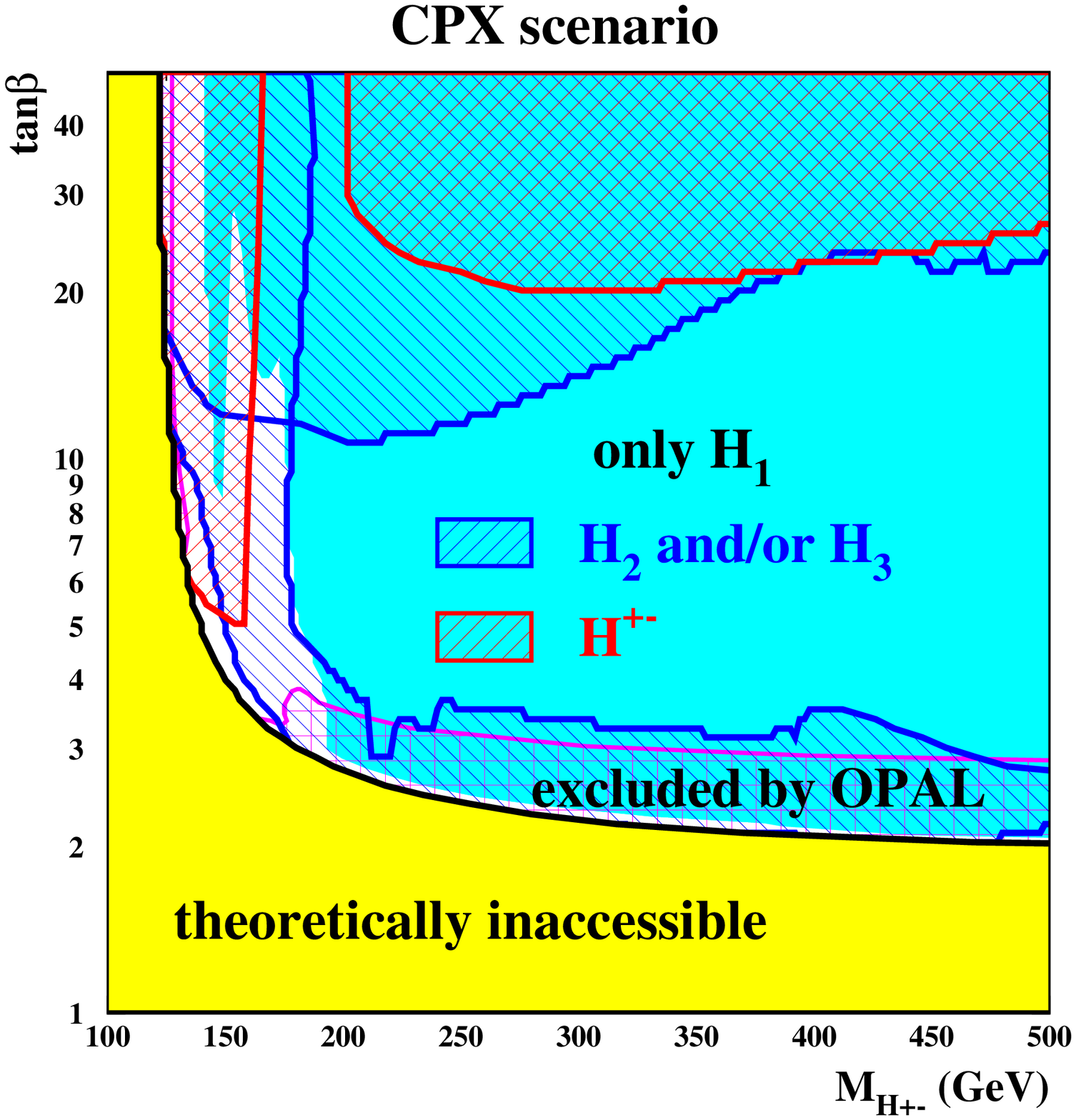,width=6.cm,height=6.cm}
\end{minipage}
&
\begin{minipage}{6.5cm}
\hspace*{-4mm}
\epsfig{file=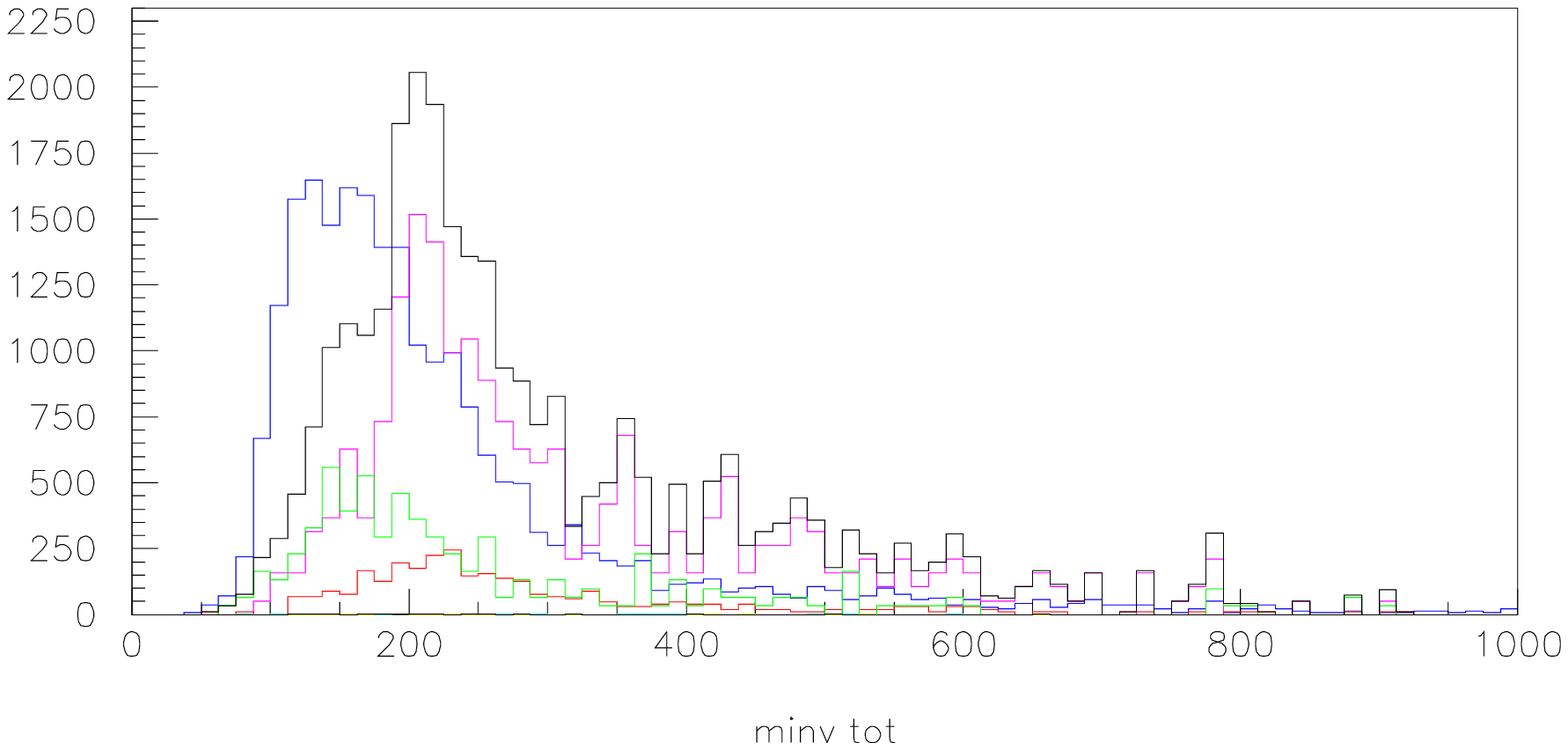,width=6.cm,height=6.6cm}
\end{minipage}
\end{tabular}
\end{center}
\vspace*{-1.8cm}
\end{figure}

\section{Conclusions and outlook}

The last few years have witnessed a large activity in the determination of the
profile of the Higgs particles in the SM and its SUSY extensions, and in their
production and detection modes at the LHC and other colliders. Major
theoretical advance has been made and in particular, the determination of the
production cross sections and the decay rates has reached a rather high level 
of accuracy; even some NNLO QCD and some electroweak corrections are now 
available. Major advance one the knowledge of the most important backgrounds 
has also been made \cite{QCD}.   

Many theoretical and experimental analyzes have been performed in the context
of the SM and MSSM and it has been shown that at least one Higgs particle
cannot escape detection at the LHC. While this fact should give us some
confidence that a breakthrough in the field will certainly occur at the LHC,
this is clearly not the end of the story and we need to perform many very 
important studies and investigate other aspects, to be ready when the machine 
starts operating. Without being exhaustive, I simply list below examples of  
points which need further efforts.  

-- We should make sure that the SM Higgs boson is observed in as many channels
as possible and in the MSSM, that the maximal number of Higgs particles is
detected.  In other words, we should work harder to extend the reach of all the
searches which have been performed up to now, and to complete them with new
ones. In the case of the MSSM for instance, we should make much smaller the
areas of parameter space in which only the lighter Higgs boson is observable
and in which several Higgs particles cannot be resolved experimentally.  

-- We should move to a more ambitious program and think more about the next
major step after discovery: how to determine experimentally the complete
profile of the Higgs particles and to unravel the mechanism of electroweak
symmetry breaking. Measuring the masses, the total widths, the couplings
to fermions and gauge bosons and the self--coupling as precisely as possible
and determining the spin and parity quantum numbers of the observed Higgs
bosons, should be a priority.  

-- We should move from the orthodox MSSM [or the $\tb$--$M_A$ parameter space
...] which has been the benchmark, besides the SM Higgs sector, that was mostly
studied up to now, and consider other more complicated or richer SM extensions. 
The SUSY regime in which some superparticles are light enough to affect the
phenomenology of the Higgs bosons, either through direct decays and production
or indirectly through loop contributions, must be scrutinized in more details. 
Extensions of the MSSM in which some basic assumptions, such as the absence of
new sources of CP--violation and/or minimal gauge group or particle content,
are relaxed should be considered more seriously. Models such as the NMSSM for
instance, which leads to a more challenging phenomenology, should be
investigated in detail.  

A number of analyzes on these issues has already been made in the recent years,
but  they need to be extended, completed and systematized. For this purpose, a
joint theoretical and experimental effort will be vital and will be required
more than it used to be in the past, as the phenomenology to be studied is
richer.  It is at this price that we will make a maximal scientific benefit from
the data to be delivered at the LHC, and to make the experiment a very
successful scientific enterprise.

\nn {\bf Acknowledgments}: I thank the organizers of the Conference, in 
particular Claudia--Elizabeth Wulz, Daniel Denegri and Giacomo Polesello, 
for the invitation to the meeting, for the very nice and warm atmosphere 
as well as for their kind patience during the writing of this contribution.

\end{document}